\documentclass[usenatbib]{mn2e}
\usepackage{epsfig}
\usepackage{times}
\def\gtrsim{~\rlap{$>$}{\lower 1.0ex\hbox{$\sim$}}}
\def\ltsim{~\rlap{$<$}{\lower 1.0ex\hbox{$\sim$}}}

\title[Dust heating in M81, M83, and NGC 2403]{Investigations of dust
heating in M81, M83, and NGC 2403 with the Herschel Space Observatory}

\author[G. J. Bendo et al.]
    {G. J. Bendo$^{1,2}$, A. Boselli$^3$, A. Dariush$^{4,5,2}$, M. Pohlen$^4$, 
         H. Roussel$^6$, M. Sauvage$^7$, \newauthor
         M. W. L. Smith$^4$, C. D. Wilson$^8$, M. Baes$^9$, A. Cooray$^{10}$,
         D. L. Clements$^2$, L. Cortese$^{11}$, \newauthor
         K. Foyle$^8$, M. Galametz$^{12}$, H. L. Gomez$^4$, 
         V. Lebouteiller$^7$, N. Lu$^{13,14}$, 
         S. C. Madden$^7$, \newauthor 
         E. Mentuch$^8$, B. O'Halloran$^2$, M. J. Page$^{15}$, A. Remy$^7$, 
         B. Schulz$^{16}$, L. Spinoglio$^{17}$\\
    $^1$  UK ALMA Regional Centre Node, Jodrell Bank Centre for Astrophysics, 
          School of Physics and Astronomy, University of Manchester, 
          Oxford Road,\\ Manchester M13 9PL, United Kingdom\\
    $^2$  Astrophysics Group, Imperial College, Blackett Laboratory,
          Prince Consort Road, London SW7 2AZ, United Kingdom\\
    $^3$  Laboratoire d' Astrophysique de Marseille, UMR6110 CNRS, 
          38 rue F. Joliot-Curie, F-13388 Marseille, France\\
    $^4$  School of Physics and Astronomy, Cardiff University, 
          Queens Buildings, The Parade, Cardiff CF24 3AA, United Kingdom\\
    $^5$  School of Astronomy, Institute for Research in Fundamental Sciences
          (IPM), PO Box 19395-5746, Tehran, Iran\\
    $^6$  Institut d'Astrophysique de Paris, UMR7095 CNRS, Universit\'e Pierre 
          \& Marie Curie, 98 bis Boulevard Arago, 75014 Paris, France\\
    $^7$  Laboratoire AIM, CEA, Universit\'e Paris Diderot, 
          IRFU/Service d'Astrophysique, Bat. 709, 91191 Gif-sur-Yvette, 
          France\\
    $^8$  Department of Physics \& Astronomy, McMaster University, Hamilton, 
          Ontario L8S 4M1, Canada\\
    $^9$  Sterrenkundig Observatorium, Universiteit Gent, Krijgslaan 281 S9,  
          B-9000 Gent, Belgium\\
    $^{10}$  Department of Physics and Astronomy, University of California, 
          Irvine, CA 92697, USA\\
    $^{11}$  European Southern Observatory, Karl-Schwarzschild Str. 2, 
          85748 Garching bei Muenchen, Germany\\
    $^{12}$  Institute of Astronomy, University of Cambridge, Madingley Road, 
          Cambridge CB3 0HA, United Kingdom\\
    $^{13}$  Jet Propulsion Laboratory, Pasadena, CA 91109, USA\\
    $^{14}$  Department of Astronomy, California Institute of Technology, 
          Pasadena, CA 91125, USA\\
    $^{15}$  Mullard Space Science Laboratory, University College London, 
          Holmbury St Mary, Dorking, Surrey RH5 6NT, United Kingdom\\
    $^{16}$  Infrared Processing and Analysis Center, California Institute of 
          Technology, Mail Code 100-22, 770 South Wilson Av, Pasadena, 
          CA 91125, USA\\
    $^{17}$  Istituto di Fisica dello Spazio Interplanetario, INAF, 
          Via del Fosso del Cavaliere 100, I-00133 Roma, Italy
}
\date{}
\pagerange{\pageref{firstpage}--\pageref{lastpage}}
\pubyear{}

\begin{document}
\label{firstpage}
\maketitle

\begin{abstract}
We use {\it Spitzer} Space Telescope and {\it Herschel} Space
Observatory far-infrared data along with ground-based optical
and near-infrared data to understand how dust heating in the nearby
face-on spiral galaxies M81, M83, and NGC 2403 is affected by the
starlight from all stars and by the radiation from star forming
regions. We find that 70/160~$\mu$m surface brightness ratios tend to
be more strongly influenced by star forming regions.  However, the
250/350~$\mu$m and 350/500~$\mu$m surface brightness ratios are more strongly
affected by the light from the total stellar populations, suggesting
that the dust emission at $>$250~$\mu$m originates predominantly from
a component that is colder than the dust seen at $<$160~$\mu$m and
that is relatively unaffected by star formation activity.  We conclude
by discussing the implications of this for modelling the spectral
energy distributions of both nearby and more distant galaxies and for
using far-infrared dust emission to trace star formation.
\end{abstract}


\section{Introduction}

Since the launch of the Infrared Astronomical Satellite \citep[IRAS;
][]{netal84}, astronomers have frequently used far-infrared emission
to trace star formation in extragalactic sources.  Far-infrared
emission has been very popular to use for estimating star formation
rates because it is not as strongly affected by dust extinction as
ultraviolet and optical tracers of star formation and because many
surveys, particularly surveys of high-redshift sources, have detected a
multitude of galaxies in the far-infrared.  Several equations have
been derived to calculate star formation rates either from
far-infrared measurements by themselves or from far-infrared data
combined with optical or ultraviolet data \citep{sy83, bx96, k98a,
ketal09}.  While it is commonly assumed in the use of these equations
that dust is primarily if not solely heated by star forming regions,
the heating source for the dust producing far-infrared emission has
been widely debated for decades.  Following the completion of the IRAS
surveys, several papers showed that IRAS 60 and 100~$\mu$m emission
was strongly correlated with star formation activity
\citep[e.g.][]{dy90, djc95, bx96}, while others argued that a
significant fraction of dust emission originated from dust heated by
evolved stars \citep[e.g][]{lh87, ws87, st92, wg96,
kcbf04}. Additionally, analyses performed by \citet{dy92} and
\citet{dy93} on IRAS data combined with 160 or 170~$\mu$m data from
the Kuiper Airborne Observatory indicated that emission at
$>$100~$\mu$m also originated from dust heated by star
formation. Later observations of nearby spiral galaxies with the {\it
Spitzer} Space Telescope \citep{wetal04} demonstrated that optical and
ultraviolet star formation tracers were very strongly correlated with
24~$\mu$m dust emission \citep{cetal05, cetal07, petal07, ketal07,
ketal09, zetal08} or with total infrared emission
\citep{ketal09}. However, similar analyses showed that the correlation
between optical/ultraviolet star formation tracers and either 70 or
160~$\mu$m emission was weaker \citep{cetal10}, which implied that not
all of the far-infrared emission originated from dust heated by star
formation. Additionally, \citet{hetal04} demonstrated that large scale
160~$\mu$m emission from M33 qualitatively appeared more similar to
large scale K-band emission than to large scale H$\alpha$, 24~$\mu$m,
or 70~$\mu$m emission, implying that the dust emitting at 160~$\mu$m
was at least partly heated by the evolved stellar population.
Furthermore, some dust models applied to IRAS, Infrared Space
Observatory \citep{ketal96}, {\it Spitzer}, and ground-based
submillimetre data for nearby galaxies have suggested that heating by
evolved stars was needed to reproduce the observed spectral energy
distributions for these galaxies \citep[e.g. ][]{detal07, dce08, b08,
ptdfkm11}.

The first results from the {\it Herschel} Space Observatory
\citep{prpetal10} brought attention back to this issue mainly because
the telescope provided data at longer wavelengths than what had been
provided by either IRAS or {\it Spitzer} and at a higher
signal-to-noise than what was possible with ground-based
instrumentation.  \citet{bwpetal10} demonstrated that colours between
160 and 500~$\mu$m in M81, an Sab galaxy \citep{detal91}, were more
strongly dependent on galactocentric radius than on far-infrared
surface brightness, which implies that the dust was heated primarily
by the evolved stars in the galaxy (which have a surface brightness
that varies primarily with radius) rather than by star forming regions
(which are primarily found in the infrared-bright regions). In
contrast, \citet{bcketal10} and \citet{vetal10} found that
100-250~$\mu$m emission from compact sources in M33, an Scd galaxy
\citep{detal91}, were strongly correlated with star formation as
measured using H$\alpha$ emission, 24~$\mu$m emission, or a
combination of the two.

The goal of this paper is to follow-up these early {\it Herschel}
results using three nearby face-on spiral galaxies with differing
Hubble types taken from the Very Nearby Galaxies Survey, a {\it
Herschel} guaranteed-time photometric and spectroscopic survey of 13
galaxies with the Photodetector Array Camera and Spectrometer
\citep[PACS;][]{pwgetal10} and Spectral and Photometric Imaging
Receiver \citep[SPIRE;][]{gaaetal10}.  We will primarily compare
surface brightness ratios between two {\it Herschel} bands to both
stellar surface brightness and to star formation rate.  Surface
brightnesses measured in single bands may be correlated to star
formation through dust heating, in which the dust is heated by the
star forming regions and becomes more luminous, or through the Schmidt
law \citep{s59, k98b}, in which star formation is correlated to the
amount of dust present.  In contrast, surface brightness ratios are
advantageous to use in that they are largely independent of the total
dust surface density.  Instead, surface brightness ratios will
primarily depend only on the average temperature of the dust emitting
at those wavelengths, with temperature variations caused either by
variations in the amount of a warmer dust component relative to a
colder dust component or variations in the overall temperature of a
single dust component.  This is similar to the approach used by
\citet{bwpetal10}, but the advantage of the analysis in this paper is
that we will not rely upon galactocentric radius as a proxy for the
surface brightness of the evolved stellar population and that we will
use a better tracer of star formation.

Section~\ref{s_sample} describes the three galaxies in our sample
(M81, M83, and NGC~2403) and why they are well-suited for this
analysis.  Section~\ref{s_data} provides an overview of the data from
{\it Herschel} and from other sources as well as details on the
preparation of the data for the analysis.  The analysis itself is
presented in Section~\ref{s_analysis}, the implications of the results
is discussed in Section~\ref{s_discussion}, and a summary is provided
in Section~\ref{s_conclusions}.

\section{Sample galaxies}
\label{s_sample}

\begin{figure*}
\begin{center}
\epsfig{file=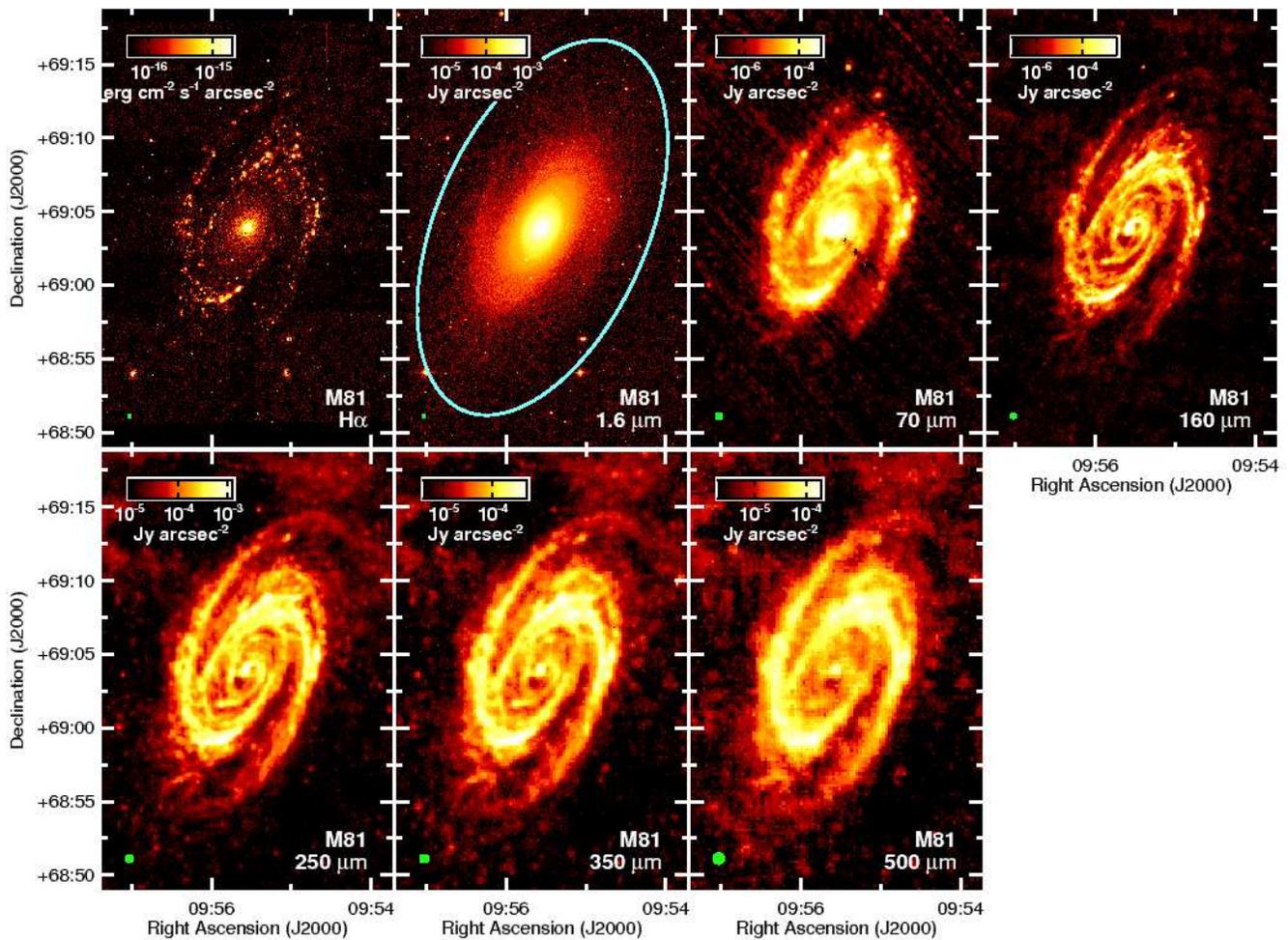}
\end{center}
\caption{The $30 \times 20$~arcmin images of M81 used in this
analysis.  The H$\alpha$ emission traces the gas that is photoionised
by star forming regions (although, in the case of M81, some residual
continuum emission from the bulge is present).  The 1.6~$\mu$m image
traces the total stellar population.  The 70-500~$\mu$m emission trace
the thermal emission from colder dust within the galaxy.  North is up
and east is left in each panel.  The green circles in the lower left
corner of each panel shows the FWHM of the data.  The cyan ellipse in
the 1.6~$\mu$m image shows the optical disk of the galaxy.}
\label{f_map_ngc3031}
\end{figure*}

\begin{figure*}
\begin{center}
\epsfig{file=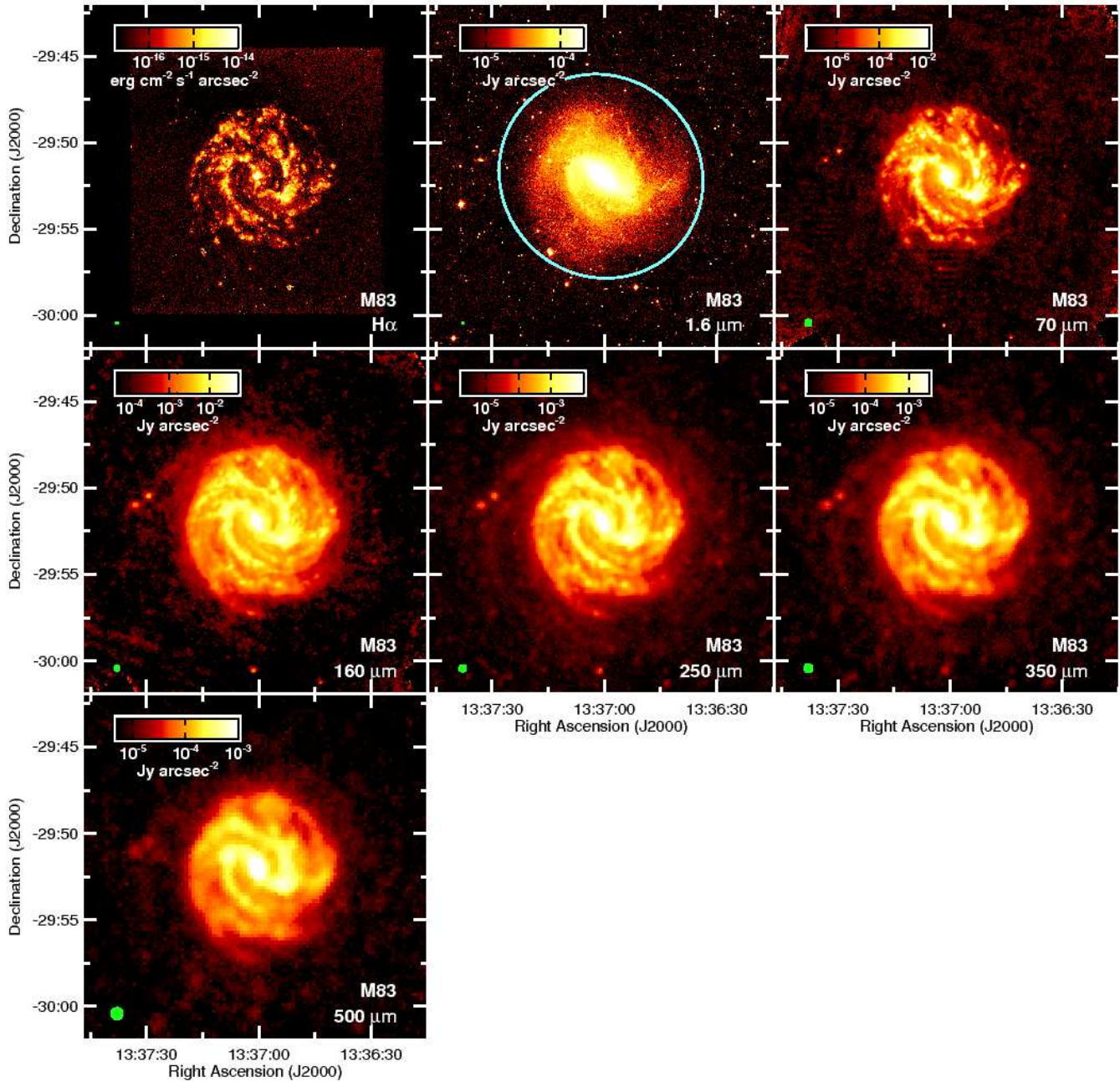}
\end{center}
\caption{The $20\times20$~arcmin images of M83 used in this analysis.
See the caption of Figure~\ref{f_map_ngc3031} for other information.}
\label{f_map_ngc5236}
\end{figure*}

\begin{figure*}
\begin{center}
\epsfig{file=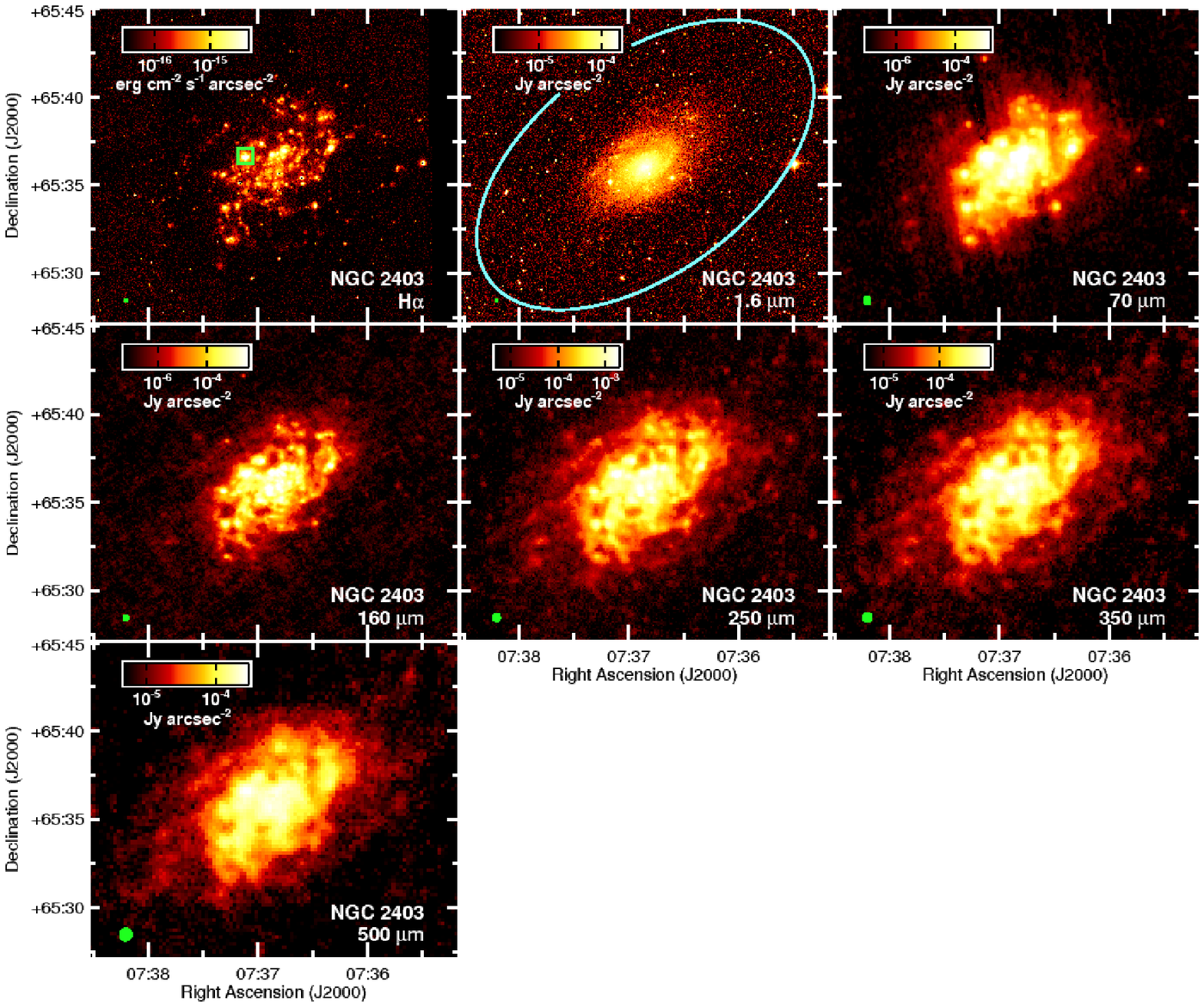}
\end{center}
\caption{The $21\times18$~arcmin images of NGC~2403 used in this
analyis.  The star forming region with the highest H$\alpha$ intensity
(VS 44) is marked with a green square.  See the caption of
Figure~\ref{f_map_ngc3031} for additional information.}
\label{f_map_ngc2403}
\end{figure*}

As stated above, the three galaxies used for this analysis were all
galaxies selected from the Very Nearby Galaxies Survey (PI: C.
Wilson), a guaranteed-time survey of 13 well-studied nearby galaxies
with a diverse range of properties.  The galaxies for the analysis in
this paper were selected because they are all spiral galaxies with
inclinations of $\ltsim60^\circ$ from face-on, they have optical discs
larger than 10~arcmin, and they are at distances of less than 5~Mpc,
so sub-kpc structures can be easily studied within the galaxies. The
galaxies are also roughly representative of early-, mid-, and late
type spiral galaxies. Details on the three galaxies used in this
analysis are given in Table~\ref{t_sample}.  The images used in the
analysis are shown in Figures~\ref{f_map_ngc3031}-\ref{f_map_ngc2403}.

\begin{table*}
\centering
\begin{minipage}{137mm}
\caption{Basic data on sample galaxies.}
\label{t_sample}
\begin{tabular}{@{}lcccccc@{}}
\hline
Galaxy &        
     Hubble &     
     Size of Optical &     
     Position Angle & 
     Inclination &         
     Distance &     
     Physical Size\\
&               
     Type $^a$ &   
     Disc (arcmin)$^b$ &   
     of Major Axis$^{bc}$ &
     &                     
     (Mpc) &       
     of 36 arcsec (pc)$^d$\\
\hline
M81 (NGC 3031) & 
     SA(s)ab & 
     $26.9 \times 14.1$ &
     $157^\circ$ &   
     $59.0^\circ$~$^e$ &     
     $3.6 \pm 0.4^f$ &
     $630 \pm 70$\\
M83 (NGC 5236) &
     SAB(s)c &
     $12.1 \times 11.5$ &
     $45^\circ$~$^g$ &
     $24^\circ$~$^g$ &
     $4.5 \pm 0.2^h$ &
     $790 \pm 30$\\
NGC 2403 &
     SAB(s)cd &
     $21.9 \times 12.3$ &
     $127^\circ$ &
     $62.9^\circ$~$^e$ &
     $3.2 \pm 0.3^f$ &
     $560 \pm 50$\\
\hline
\end{tabular}
$^a$ Hubble types are from \citet{detal91}.\\
$^b$ Data are for the $D_{25}$ isophote as given by \citet{detal91}
except where specified.\\
$^c$ Position angle is defined as degrees from north through east.\\
$^d$ This angular size represents the size of the bins used in the
analysis in this paper. See Section~\ref{s_data_convbin} for
information on the selection of this bin size.\\
$^e$ Data are taken from \citet{detal08}.\\
$^f$ Data are taken from \citet{fetal01}.\\
$^g$ Data are taken from \citet{c81}.\\
$^h$ Data are taken from \citet{tetal03}.\\
\end{minipage}
\end{table*}

M81 (NGC 3031) is an Sab galaxy \citep{detal91} that is the brightest
galaxy in the M81 Group \citep{ketal02}. As this is an early-type
spiral galaxy, the bulge is a relatively large and bright component of
the galaxy. The dust emission is unusually extended compared to other
early-type spiral galaxies; the dust distribution is similar to
late-type spiral galaxies \citep{betal07}. The dust emission also does
not qualitatively appear similar to the stellar emission.
Furthermore, the star forming regions lie primarily in the outer disc,
whereas the centre of the galaxy is relatively devoid of star
formation, as shown by the H$\alpha$ image \citep[also see ][]{djc95,
akbs97}, although incompletely-subtracted continuum emission from the
bulge is visible in the image. Consequently, this galaxy is very
useful for differentiating between effects related to heating by
starlight and effects related to heating by star forming regions.  The
galaxy does contain a low luminosity active galactic nucleus
\citep[AGN; ][]{metal08, mktdsc10} that could affect the observed
spectral energy distribution (SED) for the nuclear region.  We comment
on this in the analysis where appropriate.  Also, the 250-500~$\mu$m
images show that the galaxy is surrounded by extended emission that
was determined to be foreground cirrus emission by \citep{detal10}.
The foreground structures are fainter than much of the emission from
M81 itself, so by choosing appropriate surface brightness thresholds
for our analysis, we avoid working with data that is strongly affected
by the foreground emission.

M83 (NGC 5236) is an Sc galaxy \citep{detal91} with a grand design
appearance \citep{ee87}.  While the centre is the site of a strong
starburst \citep[e.g.][]{tetal79, th80, cetal82, betal83, tetal85,
tetal87}, star formation rates in the disc are not as extreme.  Unlike
the other two galaxies in this analysis, M83 is a case where the
stellar emission and dust emission look qualitatively
similar. However, it should be representative of many other nearby
grand-design Sb-Sc galaxies.

NGC 2403 is an Scd galaxy \citep{detal91} with a flocculent appearance
\citep{ee82, ee87}. The galaxy is the second brightest galaxy in the
M81 Group \citep{ketal02}. The extended, asymmetric dust emission in
this galaxy is typical of many Sc-Sd galaxies \citep{betal07}. The
distribution of star forming regions in NGC 2403 differs significantly
from the distribution of starlight; while the starlight peaks in the
centre of the galaxy, the regions with the strongest star forming
activity are actually seen off-centre, as seen in H$\alpha$ shown here
as well as H$\alpha$ and 24~$\mu$m images shown in previous
publications \citep[e.g.][]{detal99, betal08}.  The star forming region
with the highest H$\alpha$ and 24~$\mu$m emission, which is labeled as
region 44 (VS 44) in the catalogue of \citet{vs65} and region 128 in
the catalogue by \citet{hk83}, is located northeast of the centre of
the galaxy. Since this galaxy has multiple bright, off-centre star
forming regions, it is extremely useful for disentangling effects
dependent on star formation from effects dependent on either radius or
starlight.

\section{Observations and data reduction}
\label{s_data}

\subsection{Far-infrared and submillimetre data}

\subsubsection{MIPS 70 $\mu$m data}

Although 70~$\mu$m images were produced at the same time as 160~$\mu$m
images with PACS, we decided to use MIPS 70~$\mu$m data for M81 and
NGC~2403.  PACS data have a smaller point spread function (PSF) than
the MIPS data, so PACS data can be used to map structures on finer
scales.  As we will be matching the PSFs of the 70~$\mu$m images to
the wider PSF of the SPIRE 500~$\mu$m images, the PSF size is not
important for our analysis.  Comparisons between the PACS and MIPS
70~$\mu$m data for our sample showed that the MIPS data are more
sensitive to extended emission, which is more important for our
analysis (although both the PACS and MIPS 160~$\mu$m images have
similar sensitivities).  However, the MIPS 70~$\mu$m data suffer from
latent image effects from bright sources that manifest themselves as
dark streaks in the MIPS data.  Hence, the PACS data are preferable to
use for brighter sources, while the MIPS data are preferable to use
for fainter sources.  In the M81 and NGC~2403 data, the latent image
effects are less problematic than the lower sensitivity of the PACS
data, which is why we decided to use the MIPS data for the analysis on
those galaxies.  In images of M83, however, the PACS data have a
sufficiently high signal-to-noise that the sensitivity issues are not
as much of a concern, and the MIPS images are strongly affected by
latent image effects, so we use PACS 70~$\mu$m data for that galaxy.

The 70~$\mu$m images for M81 and NGC 2403 were acquired with the
Multiband Imaging Photometer for {\it Spitzer} \citep[MIPS;
][]{retal04} as part of the {\it Spitzer} Infrared Nearby Galaxies
Survey \citep[SINGS;][]{ketal03}. These observations consist of two
scan map observations performed using the medium scan rate (6.5 arcsec
s$^{-1}$).  The raw data from the {\it Spitzer} archive were
reprocessed using the MIPS Data Analysis Tools \citep{getal05} along
with additional processing steps. First, ramps were fit to the reads
to derive slopes.  In this step, readout jumps and cosmic ray hits
were also removed, and an electronic nonlinearity correction was
applied.  Next, the stim flash frames (frames of data in which a
calibration light source was flashed at the detectors) were used as
responsivity corrections.  After this, the dark current was subtracted
from the data, and an illumination correction was applied. Following
this, short term variations in the signal (often referred to as drift)
were removed, and additional periodic variations in the background
related to the stim flash cycle were subtracted; this also subtracted
the background from the data.  Next, a robust statistical analysis was
applied to cospatial pixels from different frames in which statistical
outliers (which could be pixels affected by cosmic rays) were masked
out.  Once this was done, final mosaics were made using pixel sizes of
4.5~arcsec.  The residual backgrounds in the data were measured in
regions outside the optical discs of the galaxies and subtracted, and
then flux calibration factors (given as 702 MJy sr$^{-1}$ [MIPS
instrumental unit]$^{-1}$ by \citet{getal07}) were applied to the
data.  An additional nonlinearity correction given as
\begin{equation}
I_{70\mu m}(\mbox{true})=0.581(I_{70\mu m}(\mbox{measured}))^{1.13}
\end{equation}
by \citet{dggetal07} was applied where the surface brightness exceeded
66 MJy sr$^{-1}$.  We also applied a colour correction of 0.901 for a
30~K blackbody given by \citet{setal07}; the actual dust temperatures
may differ from this value, but we anticipate that the associated
colour correction will be within 10\% of this value.  The final data
have a point spread function (PSF) with a full-width half-maximum
(FWHM) of 18~arcsec according to \citet{getal07}.
Calibration and nonlinearity uncertainties are estimated to be 10\%
\citep{getal07}.

\subsubsection{PACS 70-160 $\mu$m data}

The PACS observations were performed as pairs of orthogonal scans
using a 20~arcsec~s$^{-1}$ scan rate. PACS can perform simultaneous
observations in only two wave bands; the VNGS chose the 70 and 160
$\mu$m bands since they were expected to bracket the peak of the SED
better.  Four pairs of observations were performed on each galaxy.  The
observations covered areas of $40\times40$~arcmin around M81 and
NGC~2403 and $25\times25$~arcmin around M83 so as to sample an area
with a size that is at least 1.5 times the optical disc.

The PACS data were reduced with the {\it Herschel} Interactive
Processing Environment (HIPE) version 5.0 \citep{o10} and a pipeline
adapted from the official one.  After applying the standard tasks that
reformat the telemetry, associate them with the correct pointing
information, and flat-field and calibrate the signal into Jy
pixel$^{-1}$, we have applied the following specific treatments. First
we deal with the possibility for electrical crosstalk, which
potentially affects column 0 of all matrices (groups of $16\times16$
bolometers in the PACS arrays) when a bright source illuminates column
15, by systematically masking columns 0. Then we proceed to scan the
signal for glitches (cosmic rays) and other possible outliers. We use
the so-called second-level method, where what we scan are not the
individual pixel timelines but rather reconstructed timelines made of
all samples acquired on a given position of the sky. We use the {\tt
timeordered} option that is more robust to the presence of strong
brightness gradients in the sky. We check that we do not over-deglitch
the data and do not flag by mistake object structures as glitches by
projecting the glitch mask in the sky. In this glitch ``map'' the
distribution of glitches should be discontinuous and the object
structure should not be recognisable. A satisfactory setting for the
$n\sigma$ parameter of the task is 25-30 which leads to a glitch rate
of less than 1\% of the data.  We produce final maps using {\tt
Scanamorphos}\footnote[18]{Available at
http://www2.iap.fr/users/roussel/herschel/ .} (Roussel, submitted)
with its standard settings, as Scanamorphos was the best mapmaking
software available to us for removing low-frequency drift and
reproducing extended emission in PACS data.  After this, the median
residual backgrounds were measured in $\sim 1$-4~arcmin wide strips
near the edges of the observed regions and outside of the optical
discs.  These backgrounds were then subtracted from the data.

In each map, north is up, east is left.  The pixel size was set to the
default value of 1.4~arcsec at 70~$\mu$m and 2.85~arcsec at
160~$\mu$m, which corresponds to 0.25 times the FWHM of the PSF in
typical PACS observations.  The PSF has a 3-lobed structure and
pronounced Airy rings; the FWHM of the major axis in individual scan
maps in these data is 6~arcsec at 70~$\mu$m and 12~arcsec at
160~$\mu$m according to the PACS Observer's Manual from the
\citet{pacs10}\footnote[19]{The PACS Observer's Manual is available at
http://herschel.esac.esa.int/\\ Docs/PACS/pdf/pacs\_om.pdf .}.  The
calibration uncertainty is 10\% at 70~$\mu$m and 20\% at 160~$\mu$m
\citep{pwgetal10}.  Colour corrections and other effects are expected
to be small compared to the calibration uncertainty and are thus not
included here.

\subsubsection{SPIRE 250-500 $\mu$m data}

The SPIRE observations for each galaxy were performed as a pair of
orthogonal scans using a 30~arcsec~s$^{-1}$ scan rate and
nominal bias voltage settings.  The areas covered in the scan maps,
again selected to sample an area at least 1.5 times the size of the
optical disc, are $40\times40$~arcmin for M81, $20\times20$~arcmin for
M83, $30\times30$~arcmin for NGC~2403.

The SPIRE 250-500 $\mu$m observations produce timeline data that were
processed using a customized version of the official scan map pipeline
script (see \citet{getal09} and \citet{dppetal10} for more
information) run in a version of HIPE with the continuous integration
build number 4.0.1343, which was the developers' branch of the data
reduction software.  This version of the pipeline is similar to the
script in the publicly-released version 5 of HIPE, and we also used the
new flux calibration product (based on Neptune observations) that was
included in version 5.

The first steps of the timeline processing apply concurrent glitch
removal (the removal of cosmic rays that affect all detectors in an
individual array) and then wavelet glitch removal (the removal of
cosmic rays from individual detectors), with the module adjusted to
mask 7 samples following a glitch.  Next, we applied an electrical low
pass filter correction and flux calibration, and then we re-applied
the wavelet glitch removal, as, in the version of the software that we
were using, an additional deglitching step found to improve the
removal of glitches from the data.  After this, we applied a time
response correction.

We skipped applying the default temperature drift correction and
baseline subtraction and instead used a custom method called BriGAdE
(Smith et al., in preparation) to remove the temperature drift and
bring all bolometers to the same level.  For each bolometer in each
array, BriGAdE fits a linear function to the relation between the
bolometer signal timeline and the array's thermistor signal timeline
after 'jumps' (sudden DC offset in the thermistor timelines) have been
removed.  The timeline fitting is applied to the bolometer data for
the whole observation including turnaround data but excluding samples
with signals from bright sources or unidentified glitches.  For the
250 and 500~$\mu$m arrays, where we have two functional thermistors,
we used the thermistor providing the best fit to each bolometer.  The
350~$\mu$m array has only one functional thermistor, and so that
thermistor needed to be used for this processing step.  The resulting
functions, which directly relate the thermistor signal to the
background signal (including drift) measured by the bolometers, are
then used to calculate and subtract the background signal from each
bolometer signal timeline.  We have found that this method improves
the baseline subtraction significantly, especially in cases where
there are strong temperature variations during the observation.

We used the naive mapper in HIPE to create the final maps used in this
analysis.  We set the pixel size to those listed in
Table~\ref{t_spireinfo} because they are the smallest pixel sizes
possible that do not include significant numbers of map pixels that
were not crossed by detectors.  We then subtracted the median
background levels measured in $\sim4$-6 arcmin wide regions outside
the optical discs of the galaxies near the edges of the mapped area.
The SPIRE PSFs have FWHM that vary as a function of pixel scale
according to the SPIRE Observer's Manual from the
\citet{spire10}\footnote[20]{The SPIRE Observer's Manual is available
at http://\\ herschel.esac.esa.int/Docs/SPIRE/pdf/spire\_om.pdf .};
Table~\ref{t_spireinfo} lists the FWHM for the pixel sizes that we
used.  By default, the pipeline produces monochromatic flux densities
for point sources where $\nu f_\nu$ is constant, so we needed to apply
corrections that adjust the monochromatic flux density values from
point source to extended source values and colour corrections
\citep[see ][ for more information]{spire10}, both of which are also
listed in Table~\ref{t_spireinfo}.  The 350~$\mu$m data are also
multiplied by 1.0066 to correct for an issue related to an update of
the filter profile, and the other corrections are based on the updated
filter profile.  The colour corrections that we use assume that the
sources are extended and have dust SEDs that resemble blackbodies
modified with emissivity functions that scale as $\lambda^{-2}$
\citep[which is consistent with ][]{ld01}.  The temperatures are
assumed to lie between 15 and 25~K.  We compared the slopes of these
function at the central wavelengths for each SPIRE band to the slopes
in the colour correction tables given by \citet{spire10} to determine
the colour corrections given in Table~\ref{t_spireinfo}.  The final
calibration uncertainties in the data include 5\% systematic
uncertainties that is correlated across all SPIRE bands and 2\% random
uncertainties in each band \citep{spire10}.

\begin{table*}
\centering
\begin{minipage}{91mm}
\caption{Characteristics of the SPIRE 250-500~$\mu$m data and
corrections applied to the data.}
\label{t_spireinfo}
\begin{tabular}{@{}lcccc@{}}
\hline
Wave Band &      Pixel &                    PSF &      
                 Correction from Point &    Colour\\
($\mu$m) &       Scale&                     FWHM&  
                 Source to Extended &       Correction$^b$\\
&                (arcsec)&                  (arcsec)$^a$ &
                 Source Emission$^b$ &      \\
\hline
250 &            6 &                        18.2 &
                 0.9939/1.0113 &            $0.993 \pm 0.009$\\
350 &            8 &                        24.5 &
                 0.9920/1.0087 &            $1.000 \pm 0.007$\\
500 &            12 &                       36.0 &
                 0.9773/1.0065 &            $1.000 \pm 0.008$\\
\hline
\end{tabular}
$^a$ The FWHM in the map data varies with pixel scale.  These values
are for the pixel scales that we used.\\
$^b$ These are multiplicative corrections based on the values given by
the \citet{spire10}.\\
\end{minipage}
\end{table*}

\subsection{Tracer of total stellar emission}

As a tracer of the surface brightness of the total stellar population,
we used 1.6~$\mu$m (H-band) images from the ``2MASS Large Galaxy Atlas
\citep{jetal03}. These data sample the Rayleigh-Jeans side of the
thermal emission from the total stellar populations within galaxies,
and they are relatively unaffected by dust extinction.  Shorter
wavelength data would tend to be strongly affected by dust extinction
effects and could be affected more by variations in the ages of the
stellar populations.  Longer wavelength data, particularly 3.6~$\mu$m
data from {\it Spitzer}, is even less affected by dust extinction than
the 1.6~$\mu$m data.  However, multiple recent studies have
demonstrated that the 3.6~$\mu$m band may sometimes include thermal
emission from hot dust at 700-1000~K associated with sites of very
strong star formation \citep{sh09, metal09, maz10}.  \citet{maz10}
even showed that such emission has been detected in the 3.6~$\mu$m
images of M81 and NGC~2403.  Furthermore, \citet{fetal06} have shown
that polycyclic aromatic hydrocarbon spectral feature emission may
also contribute up to 50\% of the observed emission in the 3.6~$\mu$m
{\it Spitzer} band.  Hence, the 1.6~$\mu$m band appears to be the best
band to use for tracing total stellar populations, as it is the best
compromise between dust extinction effects and dust emission effects.

The processed, calibrated images were acquired from the NASA/IPAC
Extragalactic Database.  The 1.6~$\mu$m images all have pixel sizes of
1 arcsec pixel$^{-1}$.  The FWHM of the PSF is reported to be 2-3
arcsec \citep{jetal03}; we use 2.5 arcsec for our analysis.  The flux
calibration uncertainties are 3\% \citep{jetal03}. To prepare the
images for analysis, we measured median background surface
brightnesses in multiple circular regions outside the optical discs of
each galaxy and subtracted from each image.  We then identified bright
foreground stars as unresolved sources with $I_{1.6 \mu m}/I_{24 \mu
m} \gtrsim 50$ (using reprocessed 24~$\mu$m data from observations
originally performed by SINGS and by \citet{eetal05}) and removed them
by interpolating over them.

\subsection{Tracer of star formation}
\label{s_data_sf}

We use H$\alpha$ images of the galaxies as a straightforward tracer of
how star formation heats the ISM in these galaxies.  These images
trace both the photoionised gas in star forming regions themselves as
well as diffuse emission from gas that is ionised by photons escaping
from the star forming regions while tending not to include emission
from other sources.  However, the H$\alpha$ images will be affected by
extinction, with the extinction effects expected to be the strongest
in star forming regions.  This may flatten some of the relations
between H$\alpha$ intensity and other quantities.  It would be
preferable to use a tracer of star formation that corrects for dust
extinction by combining the H$\alpha$ emission (or emission in another
band) with dust emission associated with star formation, such as
24~$\mu$m emission.  However, these techniques have only been
developed for either compact sources within galaxies \citep{cetal07}
or for global flux density measurements \citep{zetal08, ketal09}.  No
corrections have been created for diffuse emission within galaxies,
and using dust emission in any band to attempt this may be difficult,
as even diffuse mid-infrared emission can potentially contain diffuse
dust emission heated by evolved stars \citep[e.g.][]{aetal98, ld01} as
well as stellar emission.  However, analyses with {\it Spitzer} data
have shown that very few H$\alpha$ sources in nearby spiral galaxies
are severely obscured by dust \citep{petal07}, and so even without
dust extinction corrections, the H$\alpha$ data should still perform
reasonably well at tracing emission from star forming regions.
Although we may expect the data to exhibit some additional scatter
without extinction corrections, the uncorrected H$\alpha$ images are
still the most suitable data for tracing how star formation regions
heat both compact and diffuse regions.

The H$\alpha$ images for M81 and NGC 2403 were originally made by
\citet{bg02} using a $1024 \times 1024$ CCD with a 0.69~arcsec
pixel$^{-1}$ scale at the 1.20~m Newton Telescope at the Observatoire
de Haute Provence. The observations for each galaxy consist of a set
of separate pointings that completely cover the optical disk of the
galaxy. The observations were performed with interferometric filters
where one was an in-band filter and the other was an off-band filter;
the H$\alpha$ image was made by subtracting one image from the other.
The FWHM of the PSF was reported to vary between 2 and 4~arcsec; we
use 3~arcsec. Calibration uncertainties in the data are 5\%.  See
\citet{bg02} for additional details.

The H$\alpha$ image for M83 was originally made by \citet{metal06} as
part of the Survey for Ionization in Neutral Gas Galaxies (SINGG) and
was distributed as part of SINGG release 1.0. The observations were
performed at the Cerro Tololo 1.5 Meter Telescope using the $2048
\times 2048$ CFCCD, which has a plate scale of
0.43~arcsec~pixel$^{-1}$. On- and off-band images were obtained using
narrowband filters; the H$\alpha$ image was made by subtracting the
on-band image from the off-band image. The reported median FWHM of the
PSF of the survey observations is 1.6~arcsec. Calibration
uncertainties are 4\%.  Additional details are provided by
\citet{metal06}.

To prepare the H$\alpha$ images for use in our analysis as a tracer of
both the compact and diffuse emission from photoionised gas, we
identified foreground stars in the image, which appeared as
incompletely subtracted point sources, and also measured and
subtracted the median background emission in circular regions at the
edges of the images (in the case of M81 and M83) or at the edge of the
region with uniform coverage (in the case of NGC~2404).  Additionally,
for M81, most of the emission within the central $5.7 \times
3.0$~arcmin (approximately the central 3~kpc) of the H$\alpha$ image
originates from either the AGN or incompletely subtracted stellar
emission from the bulge; it even has a light profile with artefacts
that are similar to the incompletely-subtracted foreground stars.
This area is masked out for the analysis.  We corrected the H$\alpha$
measurements for foreground dust extinction within the Milky Way using
calculations from the NASA/IPAC Extragalactic Database that are based
on data from \citet{sfd98}; the R-band extinction values are given in
Table~\ref{t_haforeextinct}.  We also applied corrections to remove
the [N{\small II}] line emission; we assume that the [N{\small
II}]/H$\alpha$ ratios are approximately uniform across the discs of
the galaxies.  The H$\alpha$ data from \citet{bg02} contain emission from
both the 6548 and 6583 {\AA} lines, while the data from
\citet{metal06} contain emission from only the 6583 {\AA} line.  In
the case of M81, we used data on H{\small II} regions from
\citet{gs87} to estimate the [N{\small II}] (6548 and 6583
{\AA})/H$\alpha$ ratio to be $0.40 \pm 0.13$.  For M83, the [N{\small
II}] (6583 {\AA})/H$\alpha$ ratio is estimated to be 0.34 based on the
data from \citet{betal05}.  From the radial strip data presented by
\citet{mktdsc10}, we calculated the [N{\small II}] (6548 and 6583
{\AA})/H$\alpha$ ratio for NGC~2403 to be $0.28 \pm 0.05$.

\begin{table}
\caption{Foreground extinction values used to correct H$\alpha$ fluxes.}
\label{t_haforeextinct}
\begin{center}
\begin{tabular}{@{}lc@{}}
\hline
Galaxy &     $A_R^a$\\
\hline
M81 &        0.214\\
M83 &        0.176\\
NGC 2403 &   0.077\\
\hline
\end{tabular}
\end{center}
$^a$ These values were calculated by the NASA/IPAC Extragalactic
Database using data from \citet{sfd98}.  These extinction values are
only for the foreground and do not include extinction within the target
galaxies.
\end{table}

\subsection{Convolution and binning}
\label{s_data_convbin}

To compare images from different wave bands without worrying about
beam effects, it is necessary to match the PSFs of all of the data to
the PSF of the wave band with the lowest resolution.  Otherwise, the
analysis will be strongly affected by variations related to PSF
shapes, with side lobes appearing around bright sources in maps based
on surface brightness ratios and with fainter regions appearing biased
towards emission from the wave bands with the broader PSFs.  For our
analysis, we matched the PSFs of the data to the PSF of the 500~$\mu$m
data, which has a FWHM of 36.0~arcsec in the maps.

The SPIRE PSFs are very close to Gaussian (i.e. the Airy rings 
are weak compared to the peak of the PSF), so we could use the equation
\begin{equation}
K(r)=e^{-r^2/2(\sigma_2^2-\sigma_1^2)}
\end{equation} 
to create a convolution kernel $K(r)$ that could be used to match the
PSF of data with a width $\sigma_1$ to a PSF with a width of
$\sigma_2$.  However, the MIPS 70~$\mu$m data and the PACS 70
and 160~$\mu$m data have strong Airy rings, and applying this
procedure would leave rings around bright sources.  To correct for
this, we created convolution kernels following the procedure given by
\citet{getal08} (and also see \citet{bwwetal10}).  We used
\begin{equation}
K=F^{-1} \left[ \frac{W(\omega)F[PSF_2]}{F[PSF_1]} \right]
\end{equation} 
to create convolution kernels $K$ to match PSF$_1$ to PSF$_2$.  F is a
Fourier transform and $W(\omega)$ is a radial Hanning truncation
function given by
\begin{equation}
W(\omega) = \left\{ \begin{array}{cr} 
\frac{1}{2} \left[ 1+\cos \left( \frac{2\pi\omega}{\omega_0} \right) \right]
    & \omega \leq \omega_0 \\
0 & \omega > \omega_0 \end{array} \right. 
\end{equation} 
that is used to suppress high-frequency spatial noise in the resulting
kernels by adjusting $\omega_0$.  For the MIPS 70~$\mu$m PSF, we used
radially-smoothed versions of the empirical PSFs originally presented
in \citet{ybl09}.  For the PACS PSFs, we used radially-smoothed
versions of the sum of three PSFs made from Vesta data \citep[see ][
for more information]{l10}\footnote[21]{These data are available at
ftp://ftp.sciops.esa.int/pub/hsc-calibration/\\ PACS/PSF/PACSPSF\_PICC-ME-TN-033\_v1.0.tar.gz~.~The documentation is available from
http://herschel.esac.esa.int/twiki/pub/Public/\\ PacsCalibrationWeb/bolopsfv1.01.pdf~.}. For the other PSFs, we used Gaussian functions.

\begin{figure}
\begin{center}
\epsfig{file=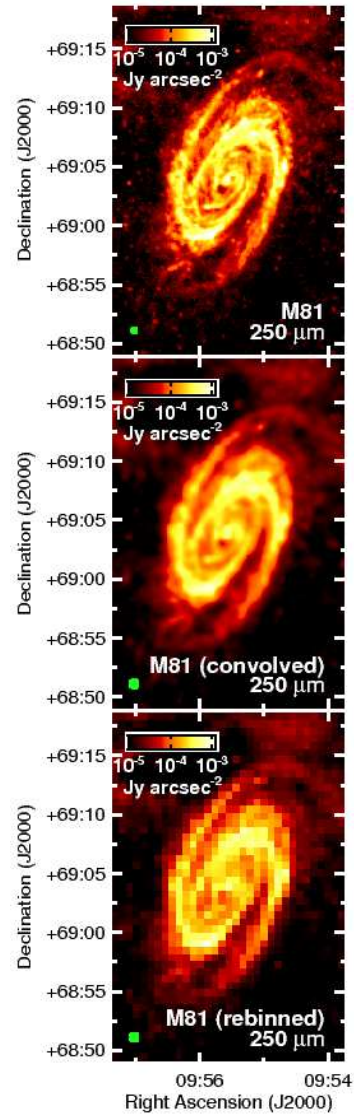}
\end{center}
\caption{The original 250~$\mu$m image of M81 (top), the 250~$\mu$m
image after it has been convolved with a kernel to match the PSF to
that of the 500~$\mu$m data (middle), and the convolved 250~$\mu$m
image after rebinning (bottom).  The green circles show the FWHM of
the PSF of the data.  In the convolved and rebinned maps, this PSF is
36~arcsec, which is also the size of the bins.  Each map is $30 \times
20$~arcmin with north up and east to the left.}
\label{f_mapbin_ngc3031}
\end{figure}

For quantitative analyses, we rebinned the data into 36~arcsec pixels
and then measured surface brightesses or intensities in all pixels
where we detected emission at the $5\sigma$ level or higher in the
70-500~$\mu$m bands.  We used this bin size because it is equivalent
to the FWHM of the PSF in the convolved data, and so we will avoid
having multiple pixels sampling the PSF of a single point-like source.
The coordinates of the bins were set up so that the centre of each
galaxy would fall within the centre of a pixel.  This binning
procedure has an advantage over either measuring surface brightnesses in
regions identified by eye or measuring surface brightnesses in regions
identified in a specific wave band in that it will be unbiased towards
any specific type of region and therefore will include both emission
from point-like sources and the diffuse ISM.  An example of the
resulting convolved and binned data is shown in
Figure~\ref{f_mapbin_ngc3031}.  In our analysis, we use measurements
in 322 bins for M81, 158 bins for M83, and 179 bins for NGC~2403.
Note that we did not use the central 36~arcsec square bin in M81 for
analysis because it includes emission from the AGN, although we do
display this data point in our figures.

\section{Analysis}
\label{s_analysis}

\subsection{Colour temperature maps} 
\label{s_analysis_images}

\begin{figure*}
\begin{center}
\epsfig{file=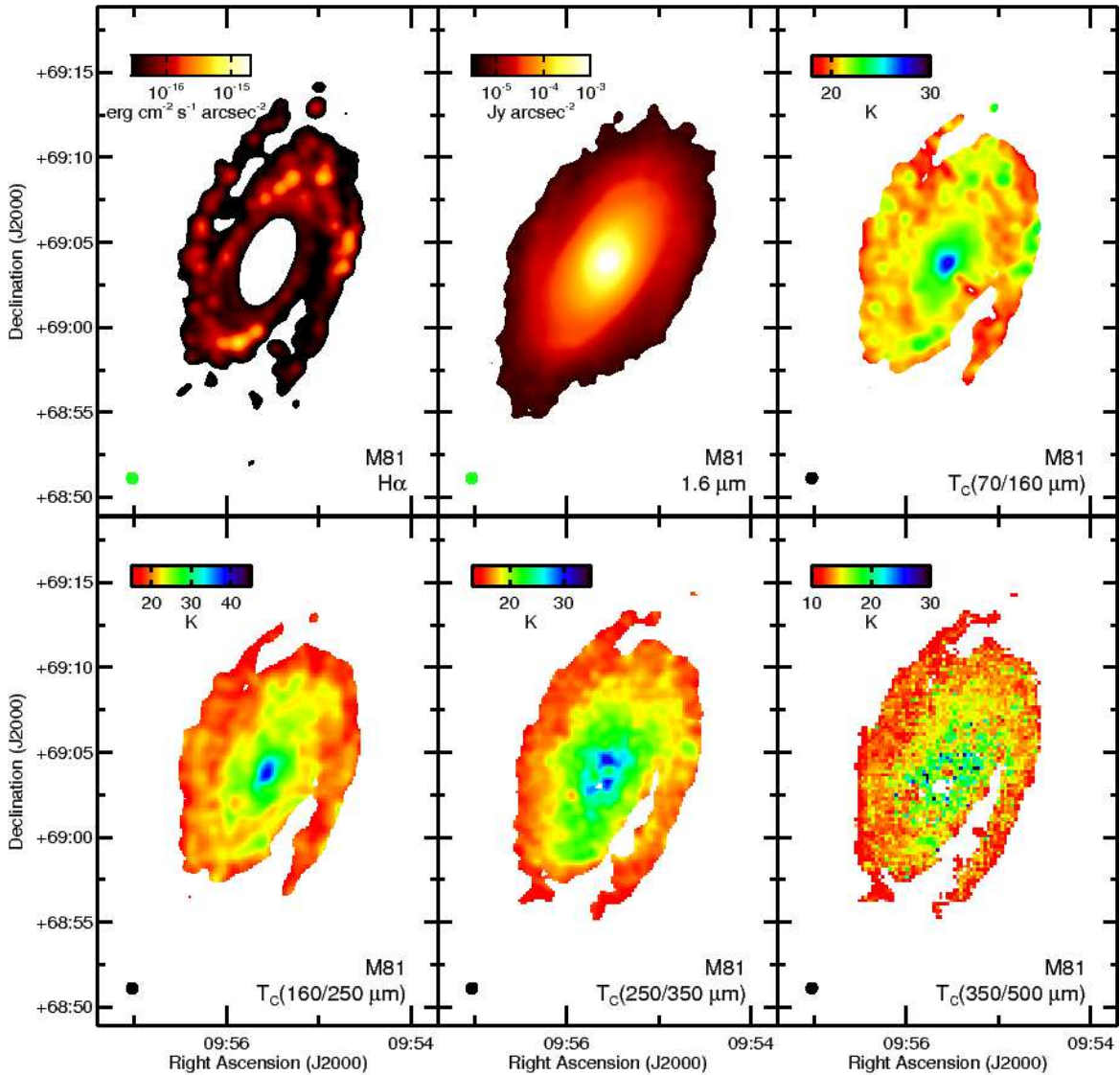}
\end{center}
\caption{The colour temperature maps of M81 and H$\alpha$ and
1.6~$\mu$m images at the same resolution.  All of these maps were
created from data where the PSF was matched to the PSF of the
500~$\mu$m data, which has a FWHM of 36~arcsec (shown by the black
circles in the lower left corner of each panel).  This 36~arcsec
circle is also the same width as the bins used in the analysis in
Sections~\ref{s_analysis_stars} and \ref{s_analysis_sf}.  Pixels
where the signal was not detected at the $5\sigma$ level in either
the single band shown (in the case of the H$\alpha$ and 1.6~$\mu$m
images) or in both bands (in the case of the colour temperature
maps) were left blank (white).  Each map is $30 \times 20$~arcmin,
and north is up and east is left in each panel.}
\label{f_maptemp_ngc3031}
\end{figure*}

\begin{figure*}
\begin{center}
\epsfig{file=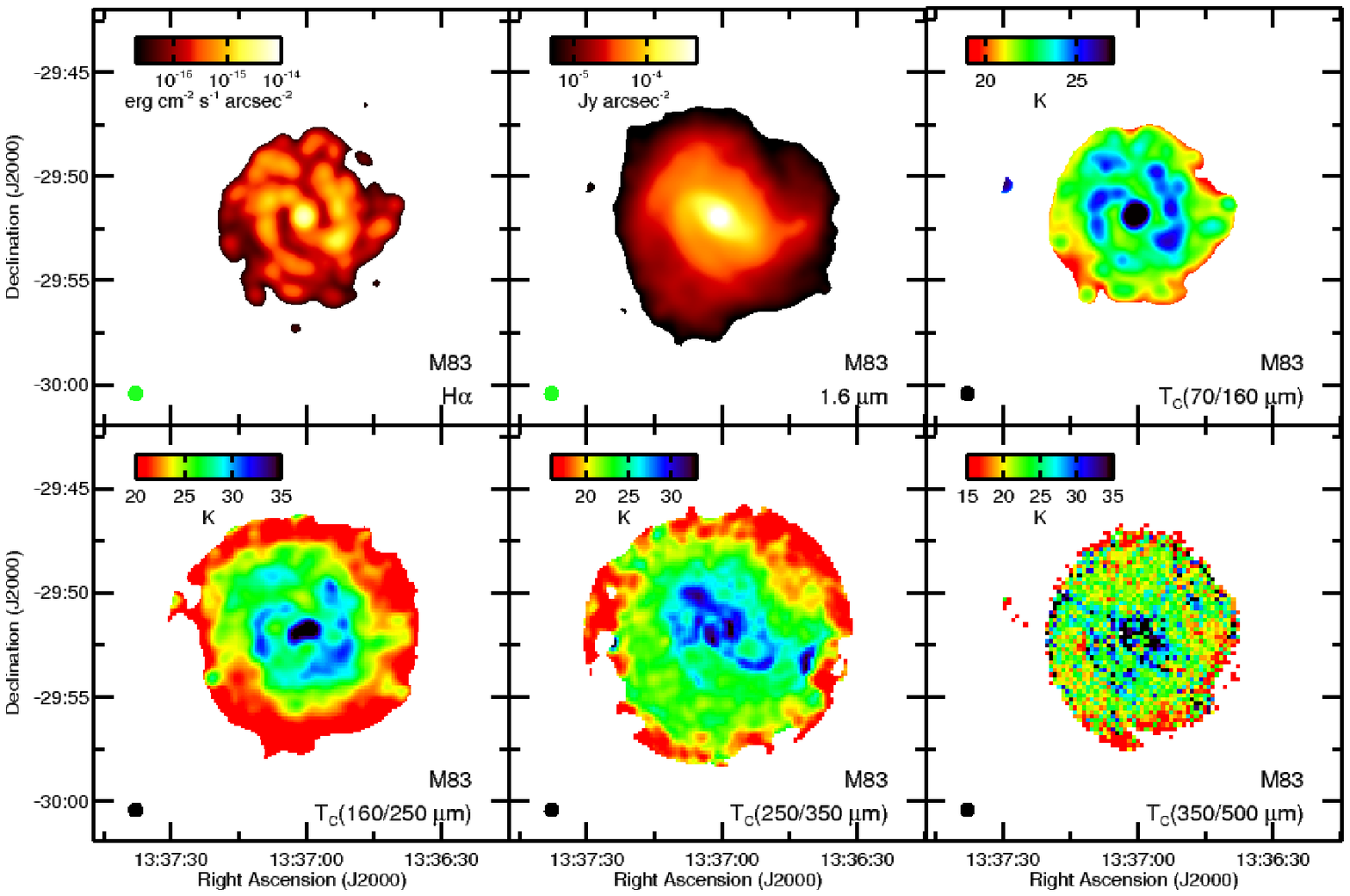}
\end{center}
\caption{The colour temperature maps of M83 and H$\alpha$ and
1.6~$\mu$m images at the same resolution.  Each map is $20\times20$~arcmin.
See the caption of Figure~\ref{f_maptemp_ngc3031} for other information
about the layout.}
\label{f_maptemp_ngc5236}
\end{figure*}

\begin{figure*}
\begin{center}
\epsfig{file=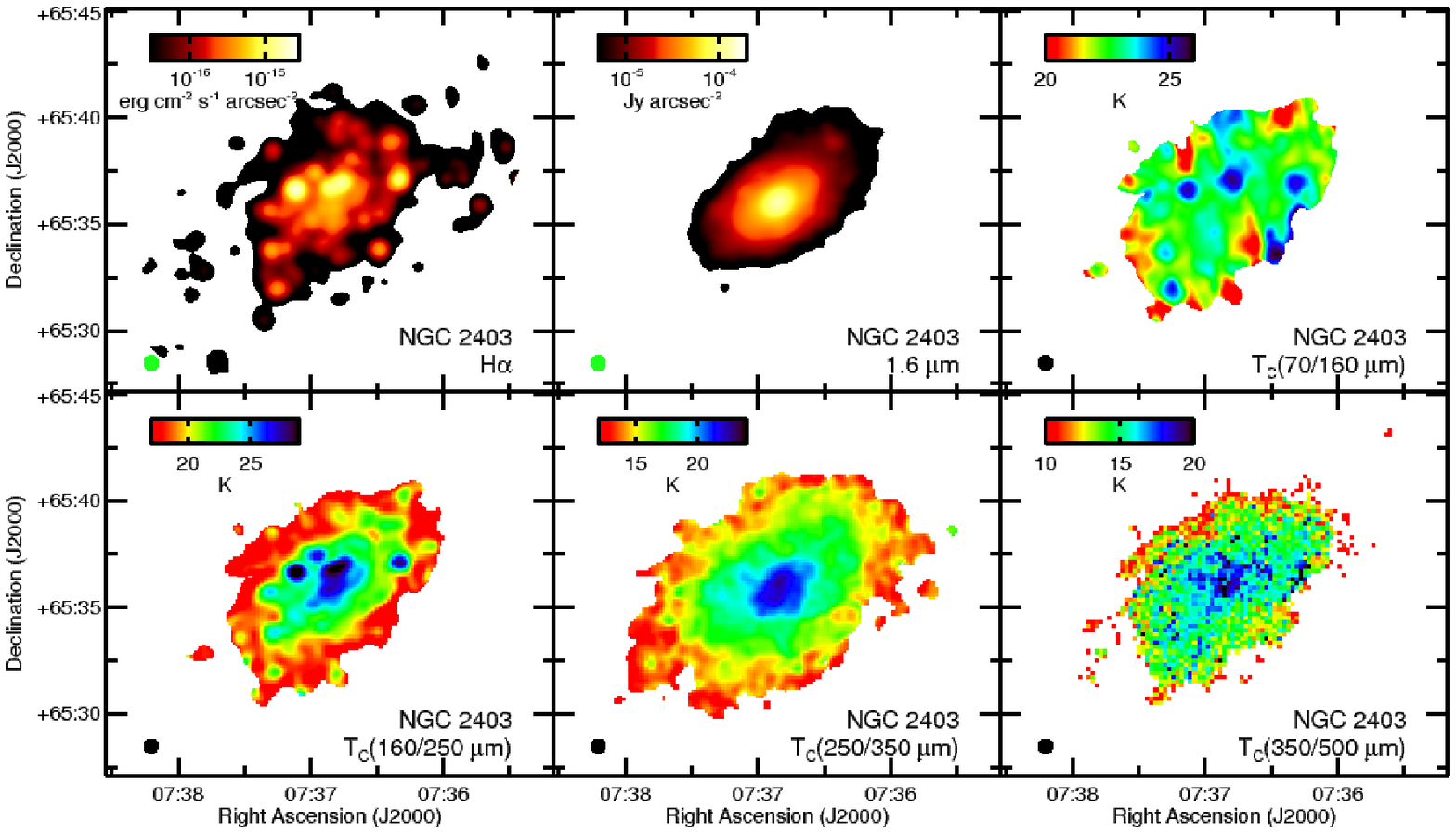}
\end{center}
\caption{The colour temperature maps of NGC~2403 and H$\alpha$ and
1.6~$\mu$m images at the same resolution.  Each map is
$21\times18$~arcmin.  See the caption of
Figure~\ref{f_maptemp_ngc3031} for other information about the
layout.}
\label{f_maptemp_ngc2403}
\end{figure*}

Figures~\ref{f_maptemp_ngc3031}-\ref{f_maptemp_ngc2403} show the
colour temperatures based on the surface brightness ratios produced by this
analysis.  The colour temperatures are calculated using 
\begin{equation}
\frac{I_{\nu}(\lambda_1)}{I_{\nu}(\lambda_2)} = \left( \frac{\lambda_2}{\lambda_1} \right) ^{3+\beta} \left( \frac{e^{hc/\lambda_2kT} - 1}{e^{hc/\lambda_1kT} - 1} \right)
\label{e_ct}
\end{equation}
in which dust emission is treated as originating from a single
component that is optically-thin in the far-infrared.  The parameter
$\beta$ is the dust emissivity coefficient that indicates how the
emissivity scales as a function of wavelength.  We use $\beta=2$
\citep[given by][]{ld01} for this calculation, as it has previously
been shown to accurately describe the observed emissivity of dust.
The colour temperature maps for each galaxy look notably different in
comparison to the stellar and dust structures in
Figures~\ref{f_map_ngc3031}-\ref{f_map_ngc2403}.  The
emission in some bands, particularly the shorter wavelength ones, may
originate from dust with a range of temperatures.  Moreover,
temperatures determined from the Rayleigh-Jeans side of the SED may be
improperly constrained without using data from the Wien side of the
peak.  Hence, these colour temperatures should be used to aid in
relating the colour variations in the maps to the SED shape and should
not necessarily be interpreted as the physical dust temperatures, even
though this could be the case for the colour temperatures based on the
longer wavelength bands.  Also, the 350/500~$\mu$m colour temperatures
tend to appear noisier.  Both wave bands sample locations on the
Rayleigh-Jeans tail of the dust SED and are therefore relatively less
sensitive to temperature variations, which causes the images to appear
noisy.

The M81 colour temperature maps are simply modified versions of the
ratio maps published by \citet{bwpetal10}.  The 70/160~$\mu$m colour
temperatures and, to a limited degree, the 160/250~$\mu$m colour
temperatures are enhanced in the spiral arms.  However, the
160/250~$\mu$m, 250/350~$\mu$m, and 350/500~$\mu$m colour temperature
maps appear to be dominated by radial variations in the colour
temperatures.  This implies that the colour variations between 160 and
500~$\mu$m are dependent either on galactocentric radius or on stellar
surface brightness.

The M83 colour temperature maps show enhancements near the nucleus and
along spiral arms near star forming regions.  The structures traced in
the 70/160~$\mu$m and 160/250~$\mu$m colour temperature maps clearly
trace H$\alpha$ regions along the edges of the spiral arms instead of
the stellar structures seen in the 1.6~$\mu$m image, suggesting that
these colour temperatures are linked to star formation activity.  We
observe these structures on the leading edges of both spiral arms,
which demonstrates that the structures are not simply the result of a
misalignment in astrometry between two pairs of images.  The
relatively smooth gradient in the 250/350~$\mu$m colour temperature is
consistent with the gradient in the 1.6~$\mu$m maps, and the central
region with 250/350~$\mu$m colour temperatures above 30K, while very
noisy, appears to correspond roughly to the bar seen in the 1.6~$\mu$m
image.  This suggests that the 250/350~$\mu$m colour temperatures
could be affected by heating by the total stellar population.  Foyle
et al. (in preparation) are performing an analysis that more carefully
examines the colour gradients across the arms and the bar.  While the
nucleus itself appears to have enhanced 70/160~$\mu$m colour
temperatures, the regions with the most strongly enhanced colour
temperatures in the other bands are the regions just to the east and
west of the nucleus.  The reason for this is unclear.  One possibility
is that the PSF matching is imperfect and residual side lobes have
appeared around the very bright nucleus in some of the data.  However,
such artefacts are not seen for other sources processed in the same
way; see, for example, VS 44 in NGC~2403.  Another possibility is that
some of the dust in the nucleus itself is relatively cold because the
dust shields itself from the nuclear starburst, but the regions
immediately outside the nucleus are not as well shielded and thus
become warmer in all wave bands.  The nuclear colour temperature
structure is poorly resolved in the resolutions that we are working at
and the structures are relatively sensitive to 1-2~arcsec adjustments
in astrometry (although the binned data are unaffected by such
adjustments), so it is difficult to accurately assess this with the
data in hand.  Future work with higher-resolution data is warranted.

The NGC~2403 colour temperature maps show significant differences in
the dust heating between 70 and 500~$\mu$m.  The 70/160~$\mu$m colour
temperature map shows temperature enhancements mainly at locations
with bright star forming regions.  However, these regions become much
less prominent in the colour temperatures based on longer wavelengths.
In the 160/250~$\mu$m and 250/350~$\mu$m colour temperature maps, the
centre of the galaxy is approximately as warm as VS 44 (the bright
star forming region to the northeast of the centre), and smooth radial
variations generally become more prominent than localised dust
temperature enhancements near star forming regions.  The
350/500~$\mu$m map appears to be similar, although the map is so noisy
that it is difficult to discern any structures.  Overall, these maps
suggest that heating by star forming regions is important between 70
and 350~$\mu$m but that variations dependent on either total stellar
surface brightness or radius becomes more important when proceeding to
wavelengths longer than 160~$\mu$m, even though the dust structures
themselves appear associated with the structures seen in H$\alpha$.

Overall, the three galaxies show similar trends.  In the colour
temperatures based on the shortest wavelength data, the temperatures
appear strongly enhanced in regions with star forming regions, but in
colour temperatures based solely on the $>$250~$\mu$m data, the
variations in the colour temperatures appear more strongly related to
either total stellar surface brightness or radius.  This suggests a
scenario in which star forming regions are located within dust clouds
that are optically thick to the blue and ultraviolet light from star
forming regions.  In this scenario, the star forming regions would
only heat a small fraction of the total dust mass in these clouds, and
this warmer dust would be the predominant emission source at
$<160$~$\mu$m.  However, most of the dust on the outside of these
clumps as well as the dust in the diffuse ISM would be shielded from
the ultraviolet and blue light from star forming regions and would
instead be heated by the diffuse interstellar radiation field.  While
this dust would be cooler, the greater mass fraction of this cooler
component would make it the more dominant emission source on the
Rayleigh-Jeans side of the dust SED at $>250$~$\mu$m.

These imaging data have shown significant qualitatitve differences
between the dust heating as a function of wavelength. In the following
sections, we will examine the different trends further using quantitative
analyses based on the binned data.

\subsection{Comparisons of surface brightness ratios to 1.6~$\mu$m emission}
\label{s_analysis_stars}

\begin{figure*}
\begin{center}
\epsfig{file=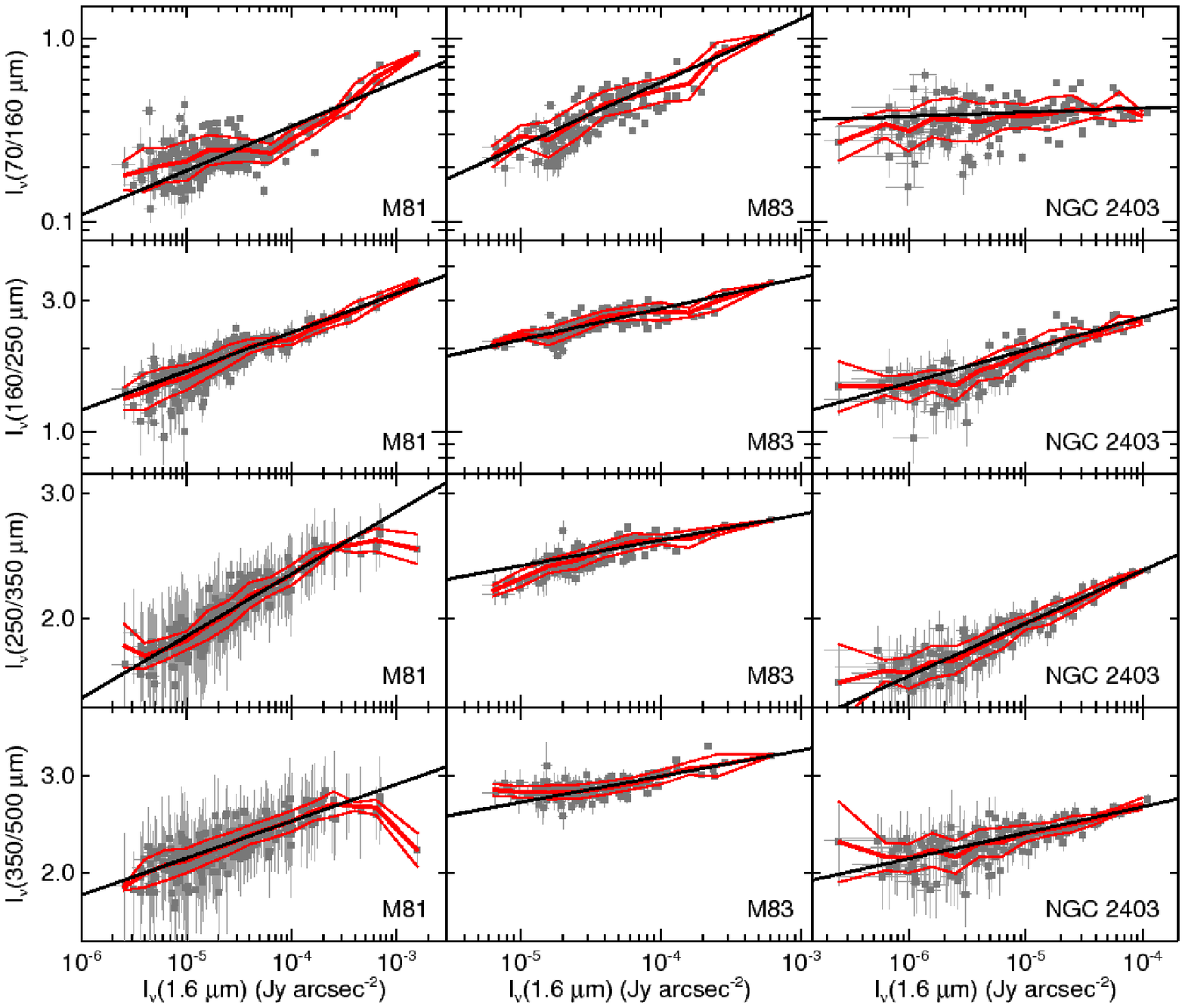}
\end{center}
\caption{Comparisons of the surface brightness ratios to the
  1.6~$\mu$m surface brightnesses for the 36 arcsec square subregions
  in the sample galaxies.  The solid lines show the best fit lines.
  The thick red line shows the weighted mean of the ratios averaged
  over intervals with a width of 0.2 in log($I_\nu$(1.6~$\mu$m)), and
  the thinner red lines show the weighted standard deviation of the
  data (except when one data point falls within an interval, in which
  case the uncertainty in the data point is used).  Parameters related
  to the fits are shown in Table~\ref{t_colorvsstar}.}
\label{f_colorvsstar}
\end{figure*}

Figure~\ref{f_colorvsstar} shows how the surface brightness ratios for
36~arcsec square subregions vary versus 1.6~$\mu$m surface brightness,
which effectively traces the starlight from the total stellar
populations within these galaxies.  Statistical information on the
relations is given in Table~\ref{t_colorvsstar}. 

\begin{table}
\caption{Results from fits between surface brightness ratios and 1.6~$\mu$m
surface brightness.}
\label{t_colorvsstar}
\begin{center}
\begin{tabular}{@{}lccc@{}}
\hline
Galaxy &     Surface &                Slope$^a$ &           Correlation\\
&            Brightness Ratio &       &                     Coefficient$^a$\\
\hline
M81 &        70/160~$\mu$m &          $0.241 \pm 0.002$ &   0.65\\
&            160/250~$\mu$m &         $0.142 \pm 0.004$ &   0.87\\
&            250/350~$\mu$m &         $0.088 \pm 0.005$ &   0.92\\
&            350/500~$\mu$m &         $0.067 \pm 0.007$ &   0.81\\
M83 &        70/160~$\mu$m &          $0.3456 \pm 0.0011$ & 0.82\\
&            160/250~$\mu$m &         $0.1137 \pm 0.0005$ & 0.79\\
&            250/350~$\mu$m &         $0.0362 \pm 0.0006$ & 0.83\\
&            350/500~$\mu$m &         $0.0474 \pm 0.0011$ & 0.64\\
NGC 2403 &   70/160~$\mu$m &          $0.023 \pm 0.002$ &   0.32\\
&            160/250~$\mu$m &         $0.120 \pm 0.002$ &   0.85\\
&            250/350~$\mu$m &         $0.074 \pm 0.002$ &   0.92\\
&            350/500~$\mu$m &         $0.047 \pm 0.003$ &   0.72\\
\hline
\end{tabular}
\end{center}
$^a$ These quantities are for the relations describing the logarithm
of the surface brightness ratios as a function of the logarithm of the
1.6~$\mu$m surface brightness in Jy arcsec$^{-2}$.  The fits are weighted
by the uncertainties in the x- and y- values.
\end{table}

Interestingly, strong correlations can be seen between the 1.6~$\mu$m
surface brightness and the 160/250~$\mu$m, 250/350~$\mu$m, and
350/500~$\mu$m ratios for all three galaxies.  In almost all cases,
the correlation coefficients are above 0.70.  The squares of the
correlation coefficients indicate the variance in one quantity that
can be related to the best fitting linear function to the other
quantity.  In this context, at least 50\% and often above 80\% of the
variance in the surface brightness ratios can be related to the
stellar surface brightness.  The 350/500~$\mu$m ratios for the three
galaxies show more scatter (i.e. have lower correlation coefficients)
than the 250/350~$\mu$m ratios in the relations with 1.6~$\mu$m
surface brightness.  This could be because of the relatively high
uncertainties in the ratios, which are in part a consequence of
sampling data on the Rayleigh-Jeans tail of the SED where temperature
variations have a relatively small effect on the colours.
Additionally, the nucleus of M81 (which corresponds to the data point
with the highest 1.6~$\mu$m surface brightness) deviates significantly
from the relations between the 1.6~$\mu$m surface brightness and
either the 250/350~$\mu$m or 350/500~$\mu$m ratios.  As stated by
\citet{bwpetal10}, this is probably because the AGN produces
non-thermal emission at submillimetre wavelengths \citep{metal08}, and
it was not included in the derivations of the statistical results
shown in Table~\ref{f_colorvsstar}.  None the less, the overall
results indicate that the total stellar populations in each galaxy
play a significant role in locally heating the dust that is emitting
at $>$160~$\mu$m.  Such a result may be expected in M81, where the
bulge has a significantly higher surface brightness in visible light
compared to the disc and where dust heating could be expected to be
dominated by evolved stars.  However, it is surprising that this
strong correlation is seen in NGC~2403, where the bulge is either very
faint or non-existent.

The results for the correlations between the 70/160~$\mu$m ratios and
the 1.6~$\mu$m surface brightness, however, show that dust heating by
the total stellar population may not be as important, although this
varies among the three galaxies examined here.  Weak correlations are
seen in M81; less than 50\% of the variance in the ratio is related to
the stellar surface brightness.  A relation may be present where the
1.6~$\mu$m surface brightness is higher than $10^{-4}$ Jy arcsec$^2$
in M81, but the relation flattens out for fainter regions.  This could
indicate that the bulge plays a role in heating the dust that produces
the 70~$\mu$m emission, but only in the centre where the bulge is very
bright.  The correlation coefficient for the relations between the
70/160~$\mu$m ratio and H$\alpha$ (discussed in
Section~\ref{s_analysis_sf}) is higher for M83, and the 70/160~$\mu$m
colour temperature map for M83 shows that the regions with enhanced
temperatures correspond to H$\alpha$ sources, which suggests that the
relation seen between the 70/160~$\mu$m ratio and 1.6~$\mu$m surface
brightness is partly a consequence of the correlation between the
stellar structures and the structures tracing star formation.  As for
NGC~2403, the correlation between the 70/160~$\mu$m ratio and
1.6~$\mu$m surface brightness is negligible, which reflects the weak
influence of the total stellar population on the colour temperatures,
as we also inferred from Figure~\ref{f_maptemp_ngc2403}.

\begin{figure*}
\begin{center}
\epsfig{file=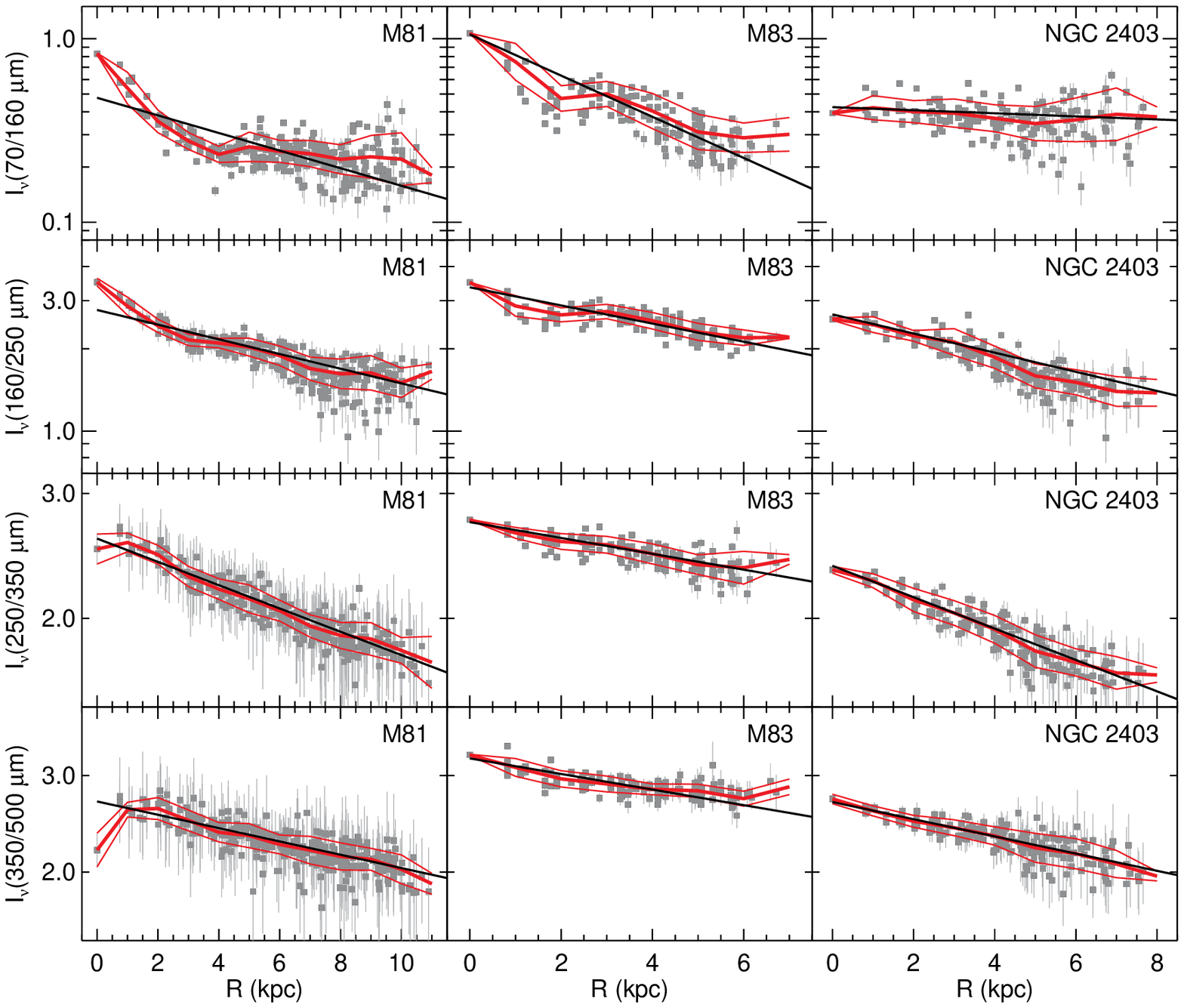}
\end{center}
\caption{Comparisons of the surface brightness ratios to deprojected
  galactocentric radius for the 36 arcsec square subregions in the
  sample galaxies.  The solid lines show the best fit lines.  The
  thick red line shows the weighted mean of the ratios averaged over
  intervals with a width of 1~kpc, and the thinner red lines show the
  weighted standard deviation of the data (except when one data point
  falls within an interval, in which case the uncertainty in the data
  point is used).  Parameters related to the fits are shown in
  Table~\ref{t_colorvsrad}.}
\label{f_colorvsrad}
\end{figure*}

\citet{bwpetal10} noted that the far-infrared surface brightness
ratios for M81 were correlated with radius.  This result was used to
infer that the ratios depended on the total stellar surface
brightness, which varies with radius within nearby galaxies.  For
completeness, we re-examined the relations for all three galaxies in
this study.  Figure~\ref{f_colorvsrad} shows how the surface
brightness ratios for the 36~arcsec square subregions vary versus
deprojected galactocentric radius, and Table~\ref{t_colorvsrad} gives
statistical information on these relations.  Visually, the relations
between radius and either the 160/250~$\mu$m, 250/350~$\mu$m, or
350/500~$\mu$m ratios appear almost as strong as the relations between
1.6~$\mu$m surface brightness and these ratios.  This is mainly
because the stellar surface brightness itself is correlated with
radius.  However, the correlation coefficients of the trends for the
relation with 1.6~$\mu$m surface brightness have slightly higher 
absolute values for all of the M81 relations, three of the relations
for M83, and two of the NGC~2403 relations.  Moreover, the
radial profiles for M81 depart from the best fit lines in the central
2~kpc in a way that would be consistent with heating by bulge stars in
this region. Furthermore, the colour temperature maps in
Figures~\ref{f_maptemp_ngc3031}-\ref{f_maptemp_ngc2403} show the
presence of some non-axisymmetric structures that correspond to spiral
arms or filaments in each of the galaxies.  Some of these features
correspond to locations with star forming regions, which can be
thought of as also enhancing the total stellar surface brightness.
Other features correspond to structures in the 1.6~$\mu$m image that
are more clearly cases where the surface brightness of the total
stellar population increases.  Hence, it would make more sense to
infer that the surface brightness ratios are more strongly dependent
on stellar surface brightness than radius.

\begin{table}
\caption{Results from fits between surface brightness ratios and
deprojected galactocentric radius.}
\label{t_colorvsrad}
\begin{center}
\begin{tabular}{@{}lccc@{}}
\hline
Galaxy &      Surface &              Slope$^a$ &            Correlation\\
&             Brightness Ratio &     &                      Coefficient$^a$\\
\hline
M81 &         70/160~$\mu$m &        $-0.0480 \pm 0.0005$ & -0.57\\
&             160/250~$\mu$m &       $-0.0268 \pm 0.0007$ & -0.82\\
&             250/350~$\mu$m &       $-0.0164 \pm 0.0009$ & -0.92\\
&             350/500~$\mu$m &       $-0.0121 \pm 0.0013$ & -0.80\\
M83 &         70/160~$\mu$m &        $-0.1122 \pm 0.0003$ & -0.77\\
&             160/250~$\mu$m &       $-0.0331 \pm 0.0001$ & -0.76\\
&             250/350~$\mu$m &       $-0.0112 \pm 0.0002$ & -0.75\\
&             350/500~$\mu$m &       $-0.0142 \pm 0.0003$ & -0.66\\
NGC 2403 &    70/160~$\mu$m &        $-0.0084 \pm 0.0007$ & -0.26\\
&             160/250~$\mu$m &       $-0.0350 \pm 0.0006$ & -0.89\\
&             250/350~$\mu$m &       $-0.0221 \pm 0.0006$ & -0.89\\
&             350/500~$\mu$m &       $-0.0157 \pm 0.0010$ & -0.80\\
\hline
\end{tabular}
\end{center}
$^a$ These quantities are for the best fit line describing the
logarithm of the surface brightness ratios as a function of
deprojected galactocentric distance in kpc.  The slope has units of
dex kpc$^{-1}$.
\end{table}

Physically, a relation between stellar surface brightness and dust
colours is more straightforward than a relation between galactocentric
radius and dust colours.  The stars can directly heat the dust, so the
stellar surface brightness would be directly linked to dust
temperature (and hence the surface brightness ratios) regardless of the
relation between stellar surface brightness and radius.  However, to
link the dust temperature with radius without connecting dust temperature
to stellar surface brightness is very difficult.  We therefore
conclude that the correlations between galactocentric radius and
either the 160/250~$\mu$m, 250/350~$\mu$m, or 350/500~$\mu$m ratio
actually reflect radial variations in the starlight heating the dust.

\begin{figure*}
\begin{center}
\epsfig{file=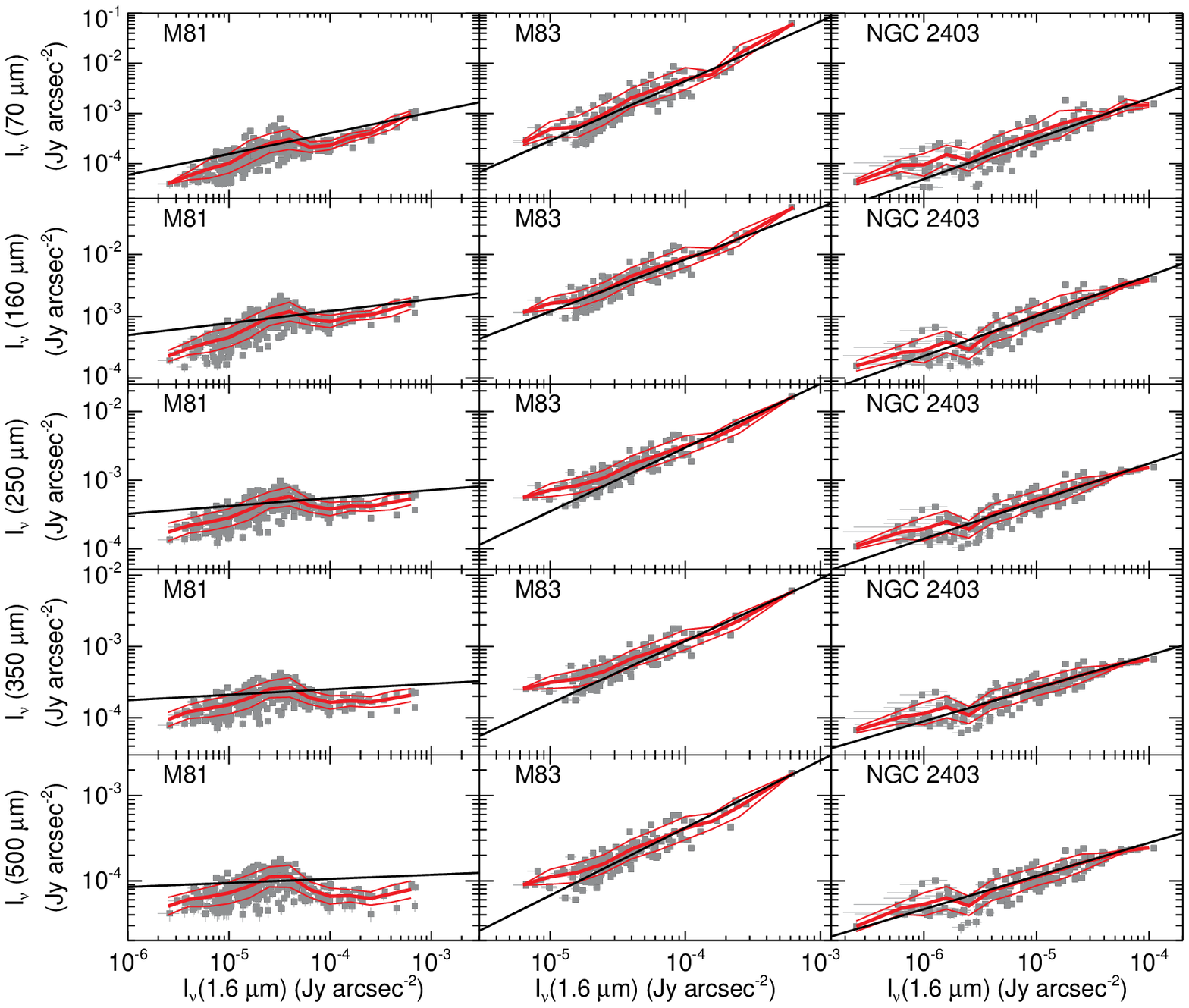}
\end{center}
\caption{Comparisons of infrared surface brightnesses in individual
  wave bands to the 1.6~$\mu$m surface brightnesses for the 36 arcsec
  square subregions in the sample galaxies.  The solid lines show the
  best fit lines.  The thick red line shows the weighted mean of the
  ratios averaged over intervals with a width of 0.2 in
  log($I_\nu$(1.6~$\mu$m)), and the thinner red lines show the
  weighted standard deviation of the data (except when one data point
  falls within an interval, in which case the uncertainty in the data
  point is used).  Parameters related to the fits are shown in
  Table~\ref{t_irvsstar}.}
\label{f_irvsstar}
\end{figure*}

Figure~\ref{f_irvsstar} shows how the 70-500~$\mu$m surface
brightnesses for the 36~arcsec square subregions are relate to the
1.6~$\mu$m surface brightnesses.  This comparison shows how the dust
emission may be related to the distribution of the total stellar
population.  Good correlations, with coefficients above 0.89,
are found between the dust and stellar emission for M83, which we
expected, as the infrared and stellar emission appear qualitatively
similar.  Surprisingly good correlations, with correlation
coefficients above 0.85, are found between dust and stellar emission
in NGC~2403 despite the signficant diffrences in the appearance of the
stellar and dust emission.  However, both the stellar and dust
emission generally decreases with galactocentric radius, which could
cause the stellar emission to appear correlated overall with the dust
emission.  In contrast to the other two galaxies, the relations
between stellar and dust emission in M81 is fairly good at 70 and
160~$\mu$m but worsens towards longer wavelengths.  Emission in the 70
and 160~$\mu$m bands is more sensitive to variations in dust temperature,
and since these bands partly trace dust within M81 that appears to be
heated by the stars seen at 1.6~$\mu$m, the 70 and 160~$\mu$m emission
is correlated with the 1.6~$\mu$m emission.  However, variations in
the 350 or 500~$\mu$m emission may be more dependent on dust surface
density than dust temperature, and so these wave bands appear more
poorly correlated with the stars in M81 because the stellar mass is
distributed differently from the dust mass.  

\begin{table}
\caption{Results from fits between far-infrared and 1.6~$\mu$m surface
    brightnesses.}
\label{t_irvsstar}
\begin{center}
\begin{tabular}{@{}lcccc@{}}
\hline
Galaxy &      Infrared Surface &    Slope$^a$ &           Correlation\\
&             Brightness&           &                     Coefficient$^a$\\
\hline
M81$^b$ &     70~$\mu$m &           $0.4164 \pm 0.0015$ & 0.78 \\
&             160~$\mu$m &          $0.1946 \pm 0.0019$ & 0.69 \\
&             250~$\mu$m &          $0.115 \pm 0.003$ &   0.52 \\
&             350~$\mu$m &          $0.077 \pm 0.004$ &   0.34 \\
&             500~$\mu$m &          $0.048 \pm 0.006$ &   0.15 \\
M83 &         70~$\mu$m &           $1.19 \pm 0.05$ &     0.90 \\
&             160~$\mu$m &          $0.84 \pm 0.03$ &     0.91 \\
&             250~$\mu$m &          $0.9284 \pm 0.0015$ & 0.90 \\
&             350~$\mu$m &          $0.8745 \pm 0.0014$ & 0.90 \\
&             500~$\mu$m &          $0.7925 \pm 0.0018$ & 0.89 \\
NGC 2403 &    70~$\mu$m &           $0.8038 \pm 0.0016$ & 0.88 \\
&             160~$\mu$m &          $0.649 \pm 0.003$ &   0.90 \\
&             250~$\mu$m &          $0.5476 \pm 0.0018$ & 0.89 \\
&             350~$\mu$m &          $0.462 \pm 0.002$ &   0.87 \\
&             500~$\mu$m &          $0.389 \pm 0.003$ &   0.87 \\
\hline
\end{tabular}
\end{center}
$^a$ These quantities are for the relations for the logarithms
of the surface brightnesses in Jy arcsec$^{-2}$.  
\end{table}

Although we have found that the distribution of dust and stellar
emission may match in many circumstances, this alone does not
necessarily indicate that the dust traced in these wave bands is
heated by the total stellar population.  Dust emission in a single
wave band is a function of both dust temperature and dust surface
density.  For example, it is possible for dust emission at 70~$\mu$m
in NGC~2403 to appear correlated with 1.6~$\mu$m emission even though
the colour temperature map and the analysis of the surface brightness
ratios suggest that the dust is not heated by the total stellar
population.  Instead, the dust and stellar surface densities are both
correlated with radius in this galaxy, causing the 70~$\mu$m emission
to appear related to 1.6~$\mu$m emission.  Conversely, it is possible
for dust emission in a single wave band, such as the 500~$\mu$m data
for M81, to appear uncorrelated with stellar surface brightness
because the wave band is more strongly affected by dust surface
density variations than dust temperature variations and because the
dust and stellar mass is distributed differently within the galaxy.
None the less, this does not preclude the possibility that the dust is
heated primarily by the total stellar population, thus causing the
temperature of the dust to be dependent upon the stellar surface
brightness.

\subsection{Comparisons of colour temperatures to star formation}
\label{s_analysis_sf}

\begin{figure*}
\begin{center}
\epsfig{file=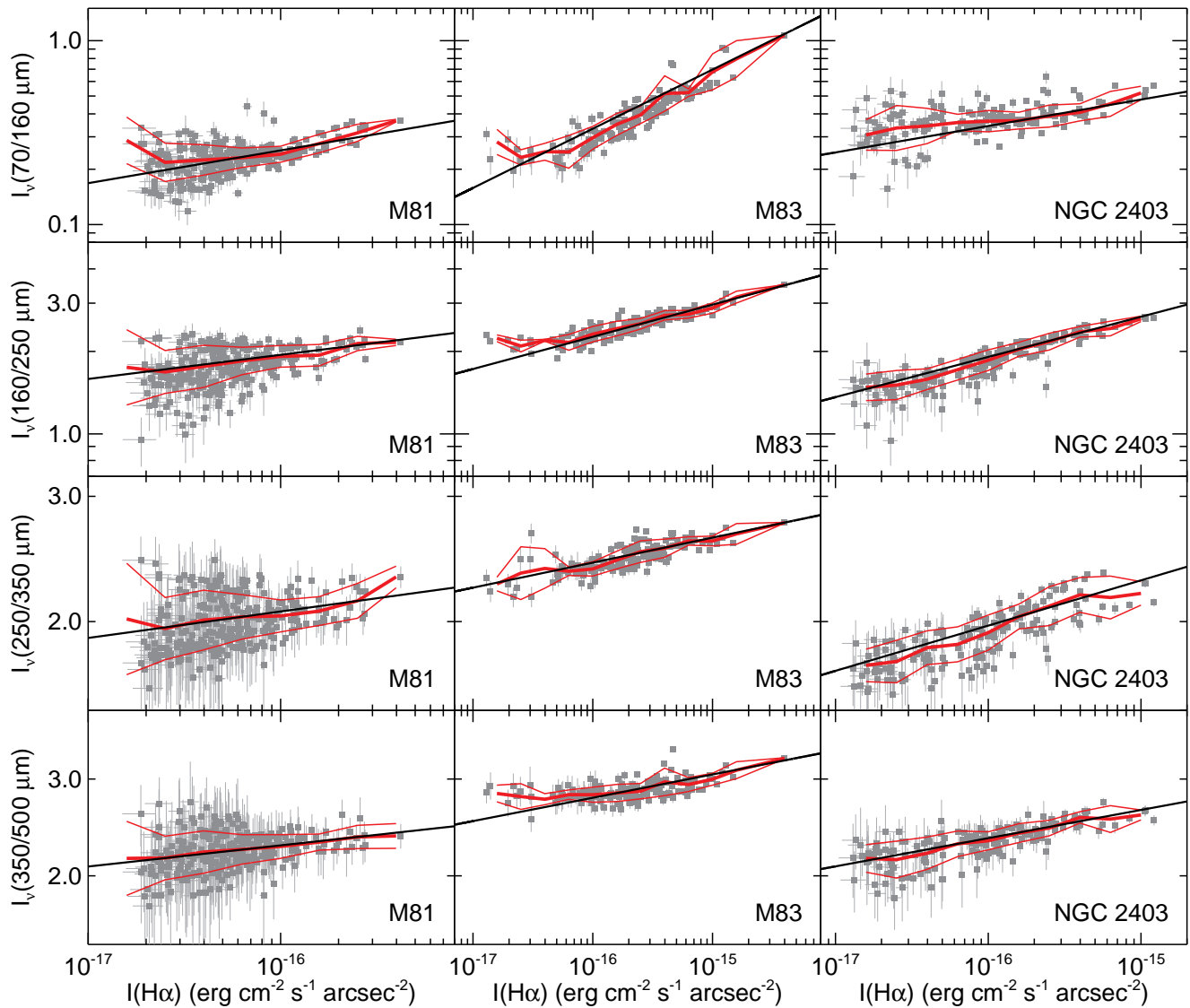}
\end{center}
\caption{Comparisons of the surface brightness ratios to the H$\alpha$
  intensities for the 36 arcsec square subregions in the sample
  galaxies.  The solid lines show the best fit lines.  The thick red
  line shows the weighted mean of the ratios averaged over intervals
  with a width of 0.2 in log($I$(H$\alpha$)), and the thinner red
  lines show the weighted standard deviation of the data (except when
  one data point falls within an interval, in which case the
  uncertainty in the data point is used).  Data from the inner central
  $5.7 \times 3.0$~arcmin (approximately the central 3~kpc) of M81 are
  not included in the analysis, as it primarily includes
  incompletely-subtracted continuum emission from the bulge.
  Parameters related to the fits are shown in
  Table~\ref{t_colorvsha}.}
\label{f_colorvsha}
\end{figure*}

Figure~\ref{f_colorvsha} shows how the colour temperatures for
36~arcsec square subregions vary versus H$\alpha$ intensity, which
traces the star formation within the galaxies.  The statistics from
these relations are given in Table~\ref{t_colorvsha}.  Each of the
galaxies in this analysis give intriguingly different results,
especially when the results are compared with those data shown in
Figure~\ref{f_colorvsstar} and Table~\ref{t_colorvsstar}.

\begin{table}
\caption{Results from fits between surface brightness ratios and H$\alpha$
intensities.}
\label{t_colorvsha}
\begin{center}
\begin{tabular}{@{}lcccc@{}}
\hline
Galaxy &      Surface &             Slope$^a$ &           Correlation\\
&             Brightness Ratio &    &                     Coefficient$^a$\\
\hline
M81$^b$ &     70/160~$\mu$m &       $0.178 \pm 0.005$ &   0.52\\
&             160/250~$\mu$m &      $0.088 \pm 0.006$ &   0.45\\
&             250/350~$\mu$m &      $0.037 \pm 0.007$ &   0.33\\
&             350/500~$\mu$m &      $0.039 \pm 0.010$ &   0.33\\
M83 &         70/160~$\mu$m &       $0.3219 \pm 0.0006$ & 0.89\\
&             160/250~$\mu$m &      $0.1174 \pm 0.0005$ & 0.87\\
&             250/350~$\mu$m &      $0.0352 \pm 0.0006$ & 0.77\\
&             350/500~$\mu$m &      $0.0425 \pm 0.0011$ & 0.50\\
NGC 2403 &    70/160~$\mu$m &       $0.144 \pm 0.002$ &   0.51\\
&             160/250~$\mu$m &      $0.147 \pm 0.003$ &   0.86\\
&             250/350~$\mu$m &      $0.064 \pm 0.002$ &   0.78\\
&             350/500~$\mu$m &      $0.051 \pm 0.003$ &   0.74\\
\hline
\end{tabular}
\end{center}
$^a$ These quantities are for the relations describing the logarithm
of the surface brightness ratios as a function of the logarithm of the
H$\alpha$ intensity in erg cm$^{-2}$ s$^{-1}$ arcsec$^{-2}$.\\ 
$^b$ Data from the inner central $5.7 \times 3.0$~arcmin
(approximately the central 3~kpc) of M81 are not included in the
analysis, as it primarily includes incompletely-subtracted continuum
emission from the bulge.
\end{table}

The data for M81 show that the surface brightness ratios are generally
poorly correlated with H$\alpha$ intensity.  This is also borne out by
the general absence of spiral structure in the colour temperature maps
in Figure~\ref{f_maptemp_ngc3031}; the colour temperatures primarily
vary with radius, and, with the exception of $T_C($70/160~$\mu$m), the
spiral structure is very weakly enhanced in the spiral arms if it is
enhanced at all.  It is possible, however, that the 70/160~$\mu$m
surface brightness ratio is enhanced in star forming regions in the
spiral arms, as the regions appear relatively bright in the colour
temperature map and as some of the data with the highest H$\alpha$
values in Figure~\ref{f_colorvsha} do appear to show a relation with
the 70/160~$\mu$m ratio.  However, the central $\sim3$~kpc of M81,
which is devoid of star formation and which has low H$\alpha$
intensities (although incompletely-subtracted continuum emission from
the bulge is seen in the H$\alpha$ image) exhibits higher
70/160~$\mu$m ratios, demonstrating that the evolved stellar
population must be the heating source for the dust emitting at
70~$\mu$m in this region.  As for the other surface brightness ratios,
the stronger correlations between the ratios and 1.6~$\mu$m surface
brightness clearly show that the total stellar population plays a much
larger role than the star forming regions in heating the dust in M81.

For M83, the correlation coefficients for the relations between
H$\alpha$ intensity and either the 70/160~$\mu$m or 160/250~$\mu$m
ratios is slightly higher than for the corresponding relations with
1.6~$\mu$m surface brightness, but the correlation coefficient for the
H$\alpha$ intensity and either the 250/350~$\mu$m or 350/500~$\mu$m
ratios is lower than the corresponding relations with 1.6~$\mu$m
surface brightness.  For the 70/160~$\mu$m, 160/250~$\mu$m, and
250/350~$\mu$m ratios, the coefficients for the relations with
H$\alpha$ intensity are within 0.10 of the coefficients for the
relations with 1.6~$\mu$m surface brightness.  This implies that both
1.6~$\mu$m and H$\alpha$ emission correlate with dust temperatures
almost equally well.  This could be true for a couple of reasons.
First of all, the H$\alpha$ and 1.6~$\mu$m emission both trace very
similar spiral structures within M83 and are correlated with each
other to some degree, especially when the data are binned up as we
have done here.  Hence, even if H$\alpha$ or 1.6~$\mu$m emission trace
completely different stellar populations, a relation found between
either band and far-infrared colours may naturally apply to the other
band as well.  It is also possible that both star forming regions and
evolved stars heat the dust observed in the far-infrared, with evolved
stars responsible for a slightly greater fraction of the heating for
dust emitting at longer wavelengths and star forming regions heating
more dust at shorter wavelengths.  A third possibility is that the
1.6~$\mu$m band includes a significant contribution from young
star-forming regions.  However, the spiral structure in the 1.6~$\mu$m
image is not as strong as the structure in the H$\alpha$ image, and
both the 1.6~$\mu$m and $T_C($250/350~$\mu$m) images exhibit bar
structures that do not have such apparent counterparts in either the
H$\alpha$, $T_C($70/160~$\mu$m), or $T_C($160/250~$\mu$m) images,
implying that ``contamination'' of the 1.6~$\mu$m band by young stars
cannot be the only explanation for the similarities in the
correlations of surface brightness ratios with either 1.6~$\mu$ or
H$\alpha$ emission.

\begin{figure}
\begin{center}
\epsfig{file=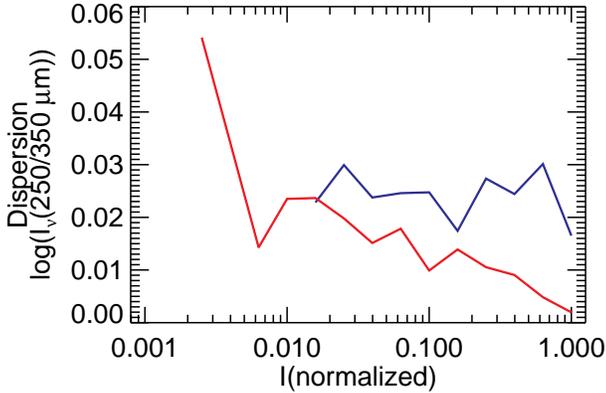}
\end{center}
\caption{The weighted standard deviation of the 250/350~$\mu$m surface
  brightness ratio for the 36 arcsec square subregions in NGC~2403
  plotted as a function of the H$\alpha$ intensity (blue) and
  1.6~$\mu$m surface brightness (red).  These curves are the same
  values used to plot the weighted standard deviations in
  Figures~\ref{f_colorvsstar} and \ref{f_colorvsha}.  The H$\alpha$
  intensity and 1.6~$\mu$m surface brightness have been normalised by
  the value of the highest surface brightness interval.}
\label{f_disp_ngc2403}
\end{figure}

The results for NGC 2403 are similar to M83 in that the 70/160~$\mu$m
are clearly more strongly correlated with H$\alpha$ intensity but the
250/350~$\mu$m surface brightness ratio is more strongly correlated
with 1.6~$\mu$m surface brightness, with the correlation coefficients
for the relations between the 160/250~$\mu$m ratio and either
H$\alpha$ intensity or 1.6~$\mu$m surface brightness appearing almost
equal.  (The correlations between the 350/500~$\mu$m ratio and either
H$\alpha$ intensity or 1.6~$\mu$m surface brightness appear almost
equally correlated probably because of the noisiness of the ratio.)
The difference in the correlation coefficients for the relations
between the 250/350~$\mu$m ratio and either 1.6~$\mu$m or H$\alpha$
emission would not by itself necessarily indicate that the total
stellar population is a significantly stronger source of heating the
dust at these wavelengths.  However, the similarities in these
correlation coefficients may be the consequence of the high scatter at
low surface brightness levels in the data.  An examination of the
regions with the highest surface brightnesses show that the
250/350~$\mu$m ratios are more strongly related to the 1.6~$\mu$m
surface brightnesses.  The 250/350~$\mu$m ratios for the binned data
in Figure~\ref{f_colorvsstar} fall within $2\sigma$ of the best
fitting relation with the 1.6~$\mu$m surface brightness for the
regions where $I_\nu(1.6$~$\mu$m$> 10^{-5}$ Jy arcsec$^{-2}$.  In
contrast, the 250/350~$\mu$m ratios for data with the highest
H$\alpha$ intensities (where$I($H$\alpha$~$> 10^{-16}$ erg cm$^{-2}$
s$^{-1}$ arcsec$^{-2}$) in Figure~\ref{f_colorvsha} frequently fall
$>3\sigma$ away from the best fit line.  This variation in the
dispersion can also be seen in Figure~\ref{f_disp_ngc2403}, which
shows the dispersion in the 250/350~$\mu$m ratio as a function of
either the H$\alpha$ or 1.6~$\mu$m emission (after normalising both to
the peak surface brightness).  Additionally, the qualitative
similarities between the $T_C($250/350~$\mu$m) and 1.6~$\mu$m maps and
the notable differences between the $T_C($250/350~$\mu$m) and
H$\alpha$ maps in Figure~\ref{f_maptemp_ngc2403} indicate that the
total stellar population is more important than star forming regions
for heating the dust observed at $>250$~$\mu$m.  While such a result
may not be surprising for M81, where the bulge is very large compared
to the disc, this is particularly surprising for NGC~2403, which is a
relatively bulgeless late-type spiral galaxy with widespread star
formation in its disc, and we anticipate that these results may apply
to other late-type spiral galaxies as well.

\begin{figure*}
\begin{center}
\epsfig{file=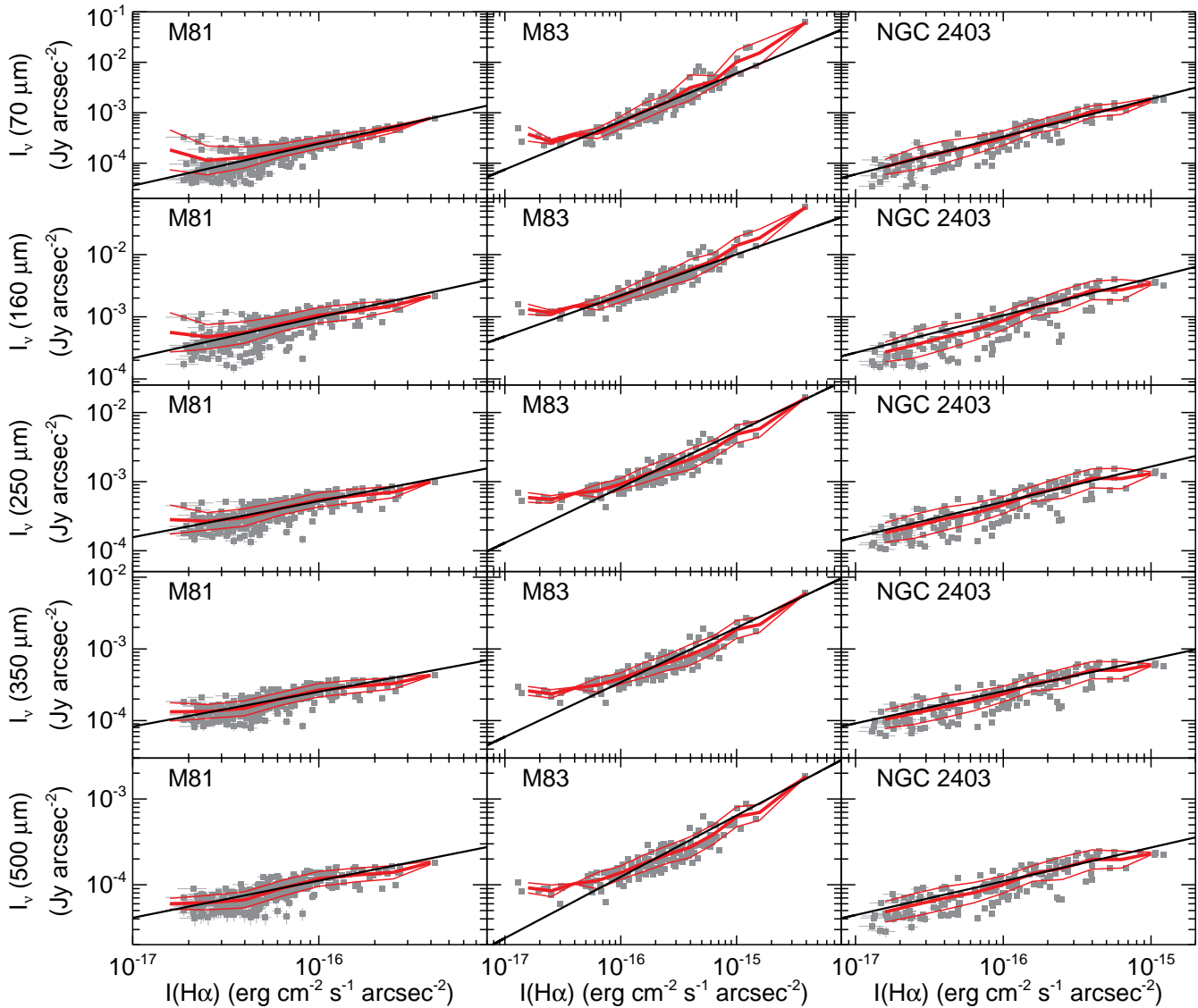}
\end{center}
\caption{Comparisons of infrared surface brightnesses in individual
  wave bands to the H$\alpha$ intensities for the 36 arcsec square
  subregions in the sample galaxies.  The solid lines show the best
  fit lines.  The thick red line shows the weighted mean of the ratios
  averaged over intervals with a width of 0.2 in log($I$(H$\alpha$)),
  and the thinner red lines show the weighted standard deviation of
  the data (except when one data point falls within an interval, in
  which case the uncertainty in the data point is used).Data from the
  inner central $5.7 \times 3.0$~arcmin (approximately the central
  3~kpc) of M81 are not included in the analysis, as it primarily
  includes incompletely-subtracted continuum emission from the bulge.
  Parameters related to the fits are shown in Table~\ref{t_irvsha}.}
\label{f_irvsha}
\end{figure*}

We also used our binned data to compare surface brightnesses from
individual far-infrared bands to H$\alpha$ emission so as to attempt
to replicate the relations between star formation tracers and 100,
160, or 250~$\mu$m emission in M33 reported by \citet{bcketal10} and
\citet{vetal10}.  Figure~\ref{f_irvsha} shows the result of this
comparison, and Table~\ref{t_irvsha} shows the slopes and correlation
coefficients for these relations.  The correlations are notably
strong.  In all cases, the correlation coefficients are above 0.7,
which indicates that at least $>50$\% of the variance in the emission
in any of these infrared bands can be accounted for by the relation
with H$\alpha$ emission.  These strong correlations are even seen in
the relations between H$\alpha$ intensity and 500~$\mu$m surface
brightness.  However, paired with the results on how the surface
brightness ratios vary with 1.6~$\mu$m surface brightness and
H$\alpha$ intensity, it is clear that the relations seen in
Figure~\ref{f_irvsha} are not necessarily a consequence of the dust
being directly heated by the star forming regions.  Instead, the
relations shown in Figure~\ref{f_irvsha} may arise as a result of the
dust being located in the same regions as the star forming regions.
We discuss the further implications of the relations between infrared
surface brightness and H$\alpha$ intensities in the next section.

\begin{table}
\caption{Results from fits between infrared surface brightnesses and H$\alpha$
intensities.}
\label{t_irvsha}
\begin{center}
\begin{tabular}{@{}lcccc@{}}
\hline
Galaxy &      Infrared Surface &    Slope$^a$ &           Correlation\\
&             Brightness&           &                     Coefficient$^a$\\
\hline
M81$^b$ &     70~$\mu$m &           $0.834 \pm 0.005$ &   0.80\\
&             160~$\mu$m &          $0.660 \pm 0.005$ &   0.74\\
&             250~$\mu$m &          $0.523 \pm 0.006$ &   0.78\\
&             350~$\mu$m &          $0.485 \pm 0.006$ &   0.79\\
&             500~$\mu$m &          $0.433 \pm 0.009$ &   0.78\\
M83 &         70~$\mu$m &           $0.95 \pm 0.03$ &     0.93\\
&             160~$\mu$m &          $0.66 \pm 0.02$ &     0.92\\
&             250~$\mu$m &          $0.8018 \pm 0.0006$ & 0.90\\
&             350~$\mu$m &          $0.7608 \pm 0.0017$ & 0.89\\
&             500~$\mu$m &          $0.7153 \pm 0.0012$ & 0.90\\
NGC 2403 &    70~$\mu$m &           $0.7474 \pm 0.0012$ & 0.91\\
&             160~$\mu$m &          $0.600 \pm 0.003$ &   0.88\\
&             250~$\mu$m &          $0.511 \pm 0.002$ &   0.86\\
&             350~$\mu$m &          $0.447 \pm 0.002$ &   0.86\\
&             500~$\mu$m &          $0.394 \pm 0.003$ &   0.86\\
\hline
\end{tabular}
\end{center}
$^a$ These quantities are for the relations describing the logarithm
of the surface brightnesses in Jy arcsec$^{-2}$ as a function of the
logarithm of the H$\alpha$ intensity in erg cm$^{-2}$ s$^{-1}$
arcsec$^{-2}$.  
\\ $^b$ Data from the inner central $5.7 \times 3.0$~arcmin
(approximately the central 3~kpc) of M81 are not included in the
analysis, as it primarily includes incompletely-subtracted continuum
emission from the bulge.
\end{table}

\subsection{Relative contributions of total stellar populations and star 
    formation to dust heating}

In prior sections, we performed comparisons between surface brightness
ratios and the emission from either the total stellar populations or
star forming regions.  When the surface brightness ratio variations
are much more strongly correlated with one source of dust heating than
with the other, we can draw very sharp conclusions about the dust
heating.  For example, we can confidently conclude that the
70/160~$\mu$m surface brightness ratio variations in NGC~2403 are
primarily dependent on heating by star forming regions, while we can
also say that the 250/350~$\mu$m variations in M81 are dependent on
heating by the total stellar population.  However, the above
comparisons have not been as effective in revealing which dust heating
source is dominant in cases where the correlations produce similar
correlation coefficients, as, for example, in the case of the
160/250~$\mu$m data for NGC~2403.

To address this issue, we attempt to fit the surface brightness ratios for
the 36~arcsec square subregions as a function of both 1.6~$\mu$m and
H$\alpha$ emission.  The equation we fit to the data is based on
\begin{equation}
CT^{4+\beta}=E_{IN},
\label{e_sblaw}
\end{equation}
a generalised version of the Stefan-Boltzmann law for the case of
blackbody emission modified by an emissivity function described by a
power law with a coefficient $\beta$.  $E_{IN}$ will be equal to the
total energy from all dust heating sources and can be written as the
sum of scaled versions of the 1.6~$\mu$m and H$\alpha$ emission.  For
a small range in temperature variation (as is the case for our data),
T can be related to a surface brightness ratio through a power law.
The difference between the actual colour temperatures and the colour
temperatures from this power law differ by only $\sim2$\% over the
range of surface brightness ratios that we are using; this is ususally
lower than the uncertainties in the ratios themselves.  Given this, we
can represent Equation \ref{e_sblaw} as
\begin{equation}
C_1 \left( \frac{I_\nu(\lambda_1)}{I_\nu(\lambda_2)} \right) ^\frac{1}{\alpha} 
    = C_2 I(\mbox{H} \alpha) + C_3 I_\nu (1.6 \mu \mbox{m}),
\end{equation}
or we can rewrite this in logarithmic form as
\begin{equation}
\ln \left( \frac{I_\nu(\lambda_1)}{I_\nu(\lambda_2)} \right) 
    = \alpha \ln(C_2 I(\mbox{H} \alpha) + C_3 I_\nu (1.6 \mu \mbox{m}))
    - \ln(C_1).
\label{e_efit}
\end{equation}
Equation \ref{e_efit} can be simplified as
\begin{equation}
\ln \left( \frac{I_\nu(\lambda_1)}{I_\nu(\lambda_2)} \right)
    = \alpha \ln(I(\mbox{H} \alpha) + A_1 I_\nu (1.6 \mu \mbox{m}))
    + A_2.
\label{e_efinal}
\end{equation}
Equation~\ref{e_efinal} can be thought of as similar to the relations
typically shown between colour and flux or variants of these
quantities (such as Figures~\ref{f_colorvsstar} and \ref{f_colorvsha})
but with two flux quantities simultaneously fit to the colour instead
of one.  The relative contribution of star forming regions to the dust
heating can be calculated using
\begin{equation}
\frac{E_{SF}}{E_{Total}} = \frac{I(\mbox{H} \alpha)}
    {I(\mbox{H} \alpha) + A_1 I_\nu (1.6 \mu \mbox{m})}.
\label{e_efrac}
\end{equation}
This approach to determining the relative contributions of star
forming regions and the total stellar population to dust heating is
relatively simplistic.  More complex models would be able to more
accurately describe the propagation of photons through the ISM, the
absorption of light by dust, the range of dust temperatures, and the
dust emission, but this requires making many assumptions about the
stellar populations and the physical properties of the dust particles.
In contrast, the analysis we have outlined above relies primarily upon
the assumptions that dust heating occurs locally and that the relation
between surface brightness ratios and temperatures follow power laws.
Hence, it can be used as a way to examine dust heating from two energy
sources using measurements that are independent of dust surface
density in a way that relies upon very few assumptions.

\begin{figure*}
\begin{center}
\epsfig{file=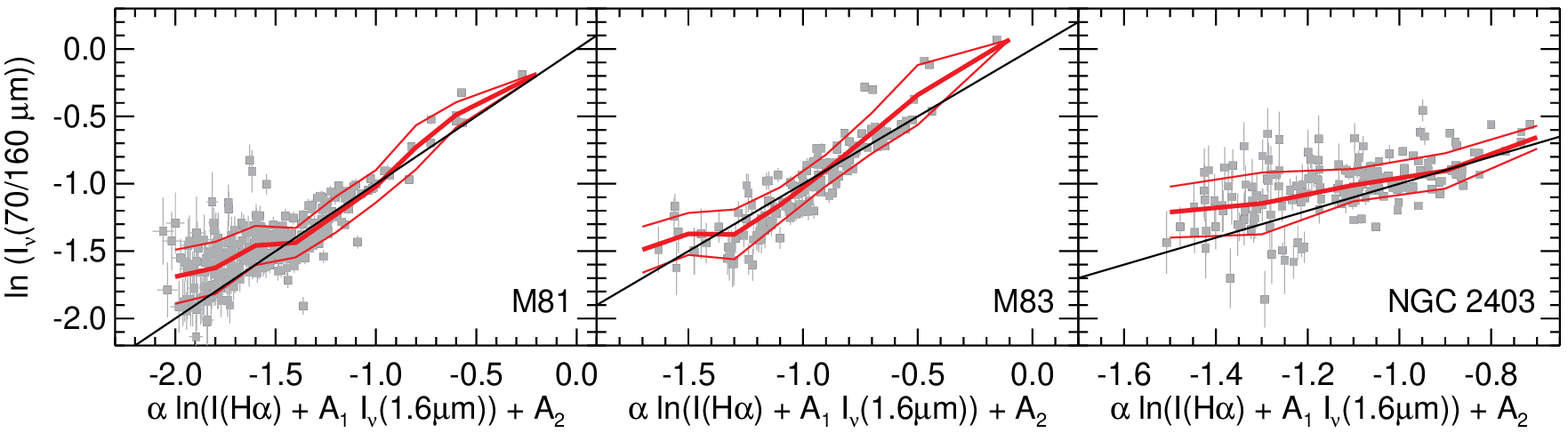}
\epsfig{file=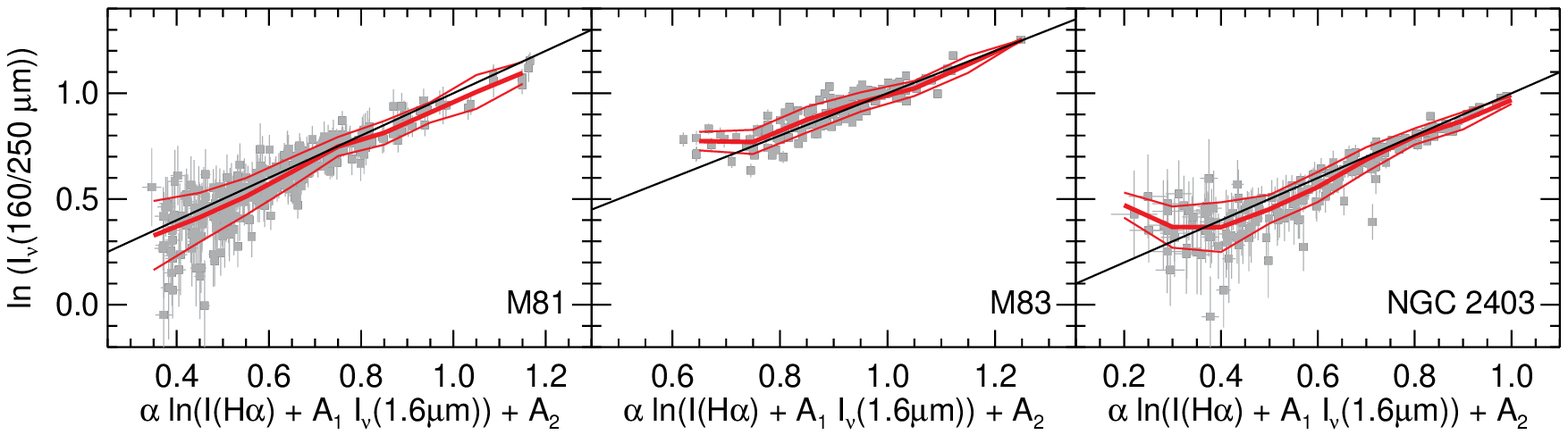}
\epsfig{file=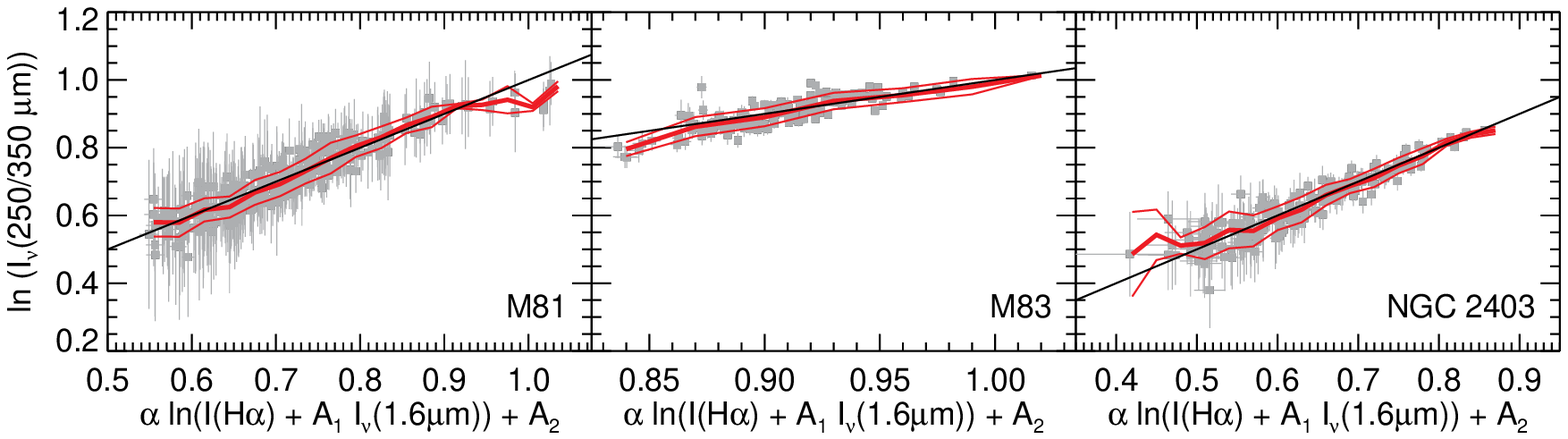}
\epsfig{file=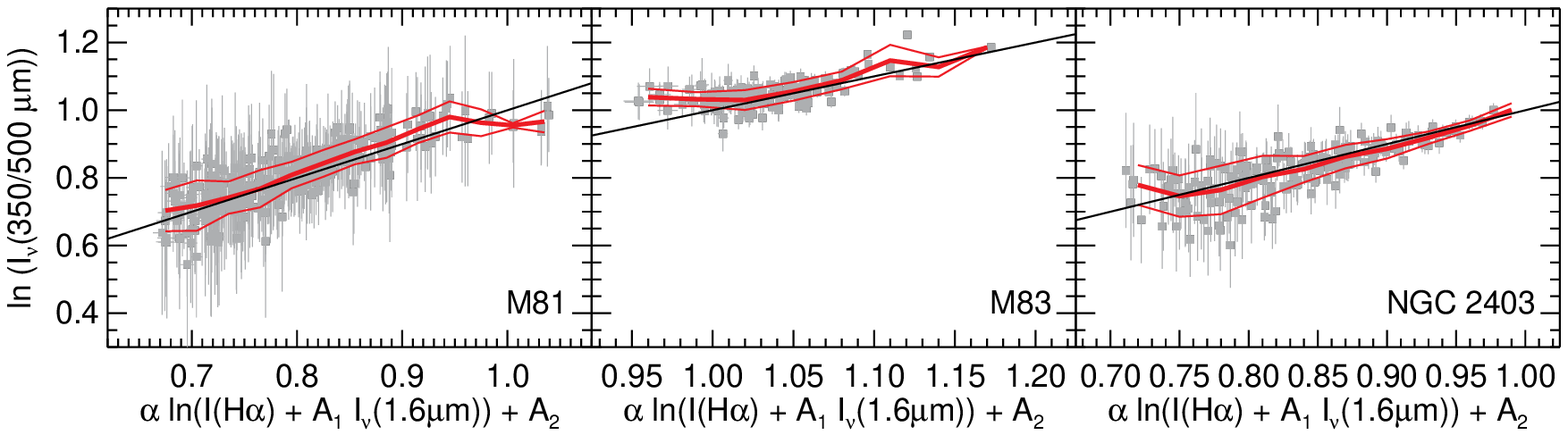}
\end{center}
\caption{Comparisons of the surface brightness ratios to the relation
  described by Equation~\ref{e_efinal} for the 36 arcsec square
  subregions in the sample galaxies.  The solid lines show the best
  fit lines, which have a slope of 1.  The thick red line shows the
  weighted mean of the ratios averaged over different interval widths
  for each surface brightness ratio (0.2 in the units of the x-axes
  for the 70/160~$\mu$m ratio, 0.1 for the 160~$\mu$m ratio, and 0.03
  for the 250/350~$\mu$m and 350/500~$\mu$m ratios), and the thinner
  red lines show the weighted standard deviation of the data (except
  when one data point falls within an interval, in which case the
  uncertainty in the data point is used).  Data from the inner central
  $3 \times 3$ subregions ($108 \times 108$~arcsec) of M83 are not
  included in the analysis, as it caused problems during the fit as
  described in the text.  Parameters related to the fits are shown in
  Table~\ref{t_colorvsmult}.}
\label{f_multfit}
\end{figure*}

\begin{figure*}
\begin{center}
\epsfig{file=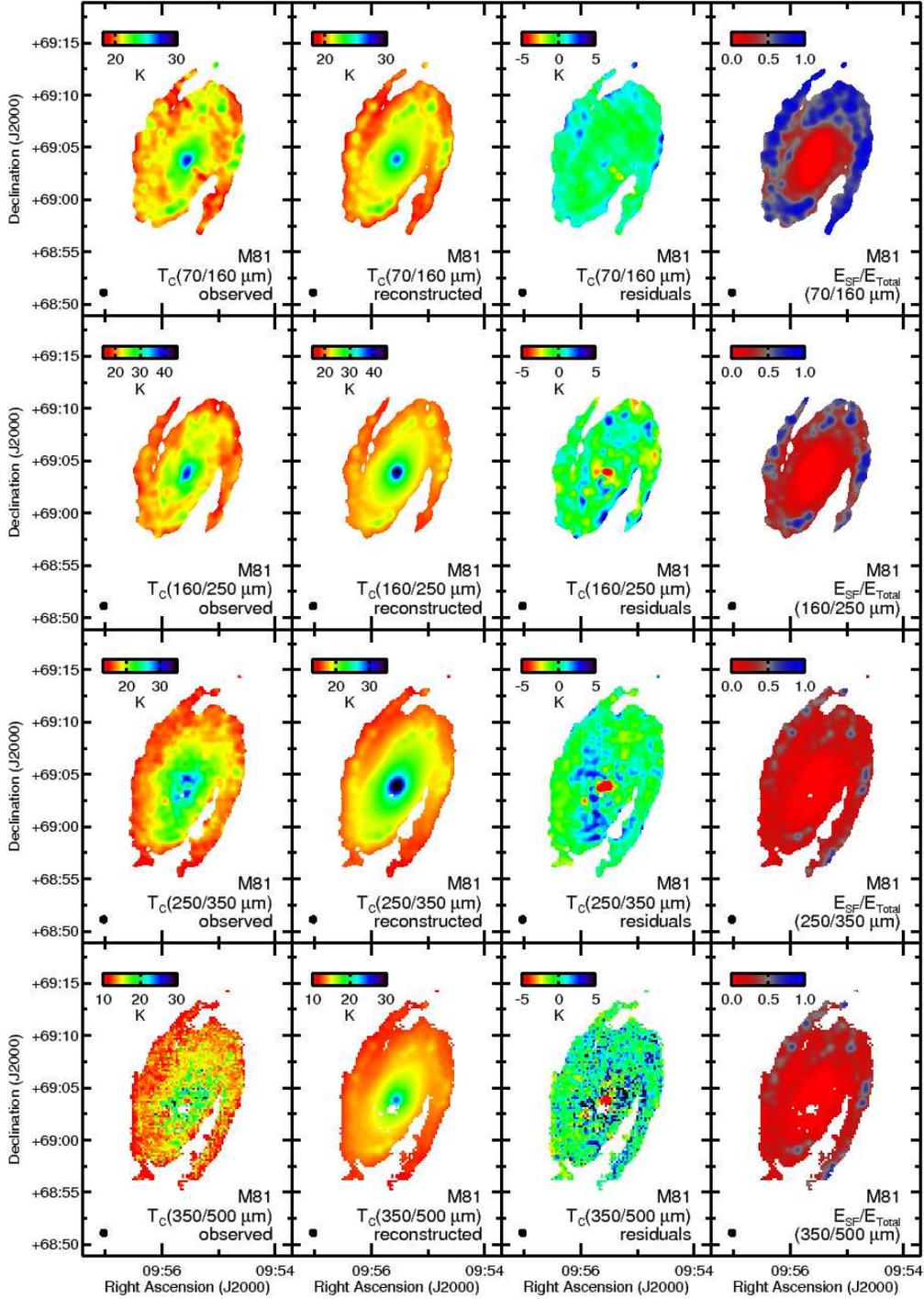,height=20cm}
\end{center}
\caption{The left column shows the observed colour temperature maps of
  M81.  The next column shows the colour temperature maps
  reconstructed using Equation~\ref{e_ct} and \ref{e_efinal}, the
  H$\alpha$ and 1.6~$\mu$m images, and the parameters if
  Table~\ref{t_colorvsmult}.  The third column shows the residual
  values when the reconstructed image is subtracted from the observed
  image.  The fourth column shows the fraction of dust heating that is
  attributed to star formation as traced by the given surface
  brightness ratio; these images are created using
  Equation~\ref{e_efrac}, the H$\alpha$ and 1.6~$\mu$m images, and the
  $A_1$ parameters in Table~\ref{t_colorvsmult}. All of these maps
  were created from data where the PSF was matched to the PSF of the
  500~$\mu$m data, which has a FWHM of 36~arcsec (shown by the black
  circles in the lower left corner of each panel).  This 36~arcsec
  circle is also the same width as the bins used in the analysis in
  Sections~\ref{s_analysis_stars} and \ref{s_analysis_sf}.  Pixels
  where the signal was not detected at the $5\sigma$ level in both
  bands used to create the observed colour temperature maps were left
  blank (white).  In the other three columns, these pixels are left
  blank and pixels where data are not detected at the $5\sigma$ level
  in either the H$\alpha$ or 1.6~$\mu$m images were left blank.  Each
  map is $30 \times 20$~arcmin, and north is up and east is left in
  each panel.}
\label{f_mapfrac_ngc3031}
\end{figure*}

\begin{figure*}
\begin{center}
\epsfig{file=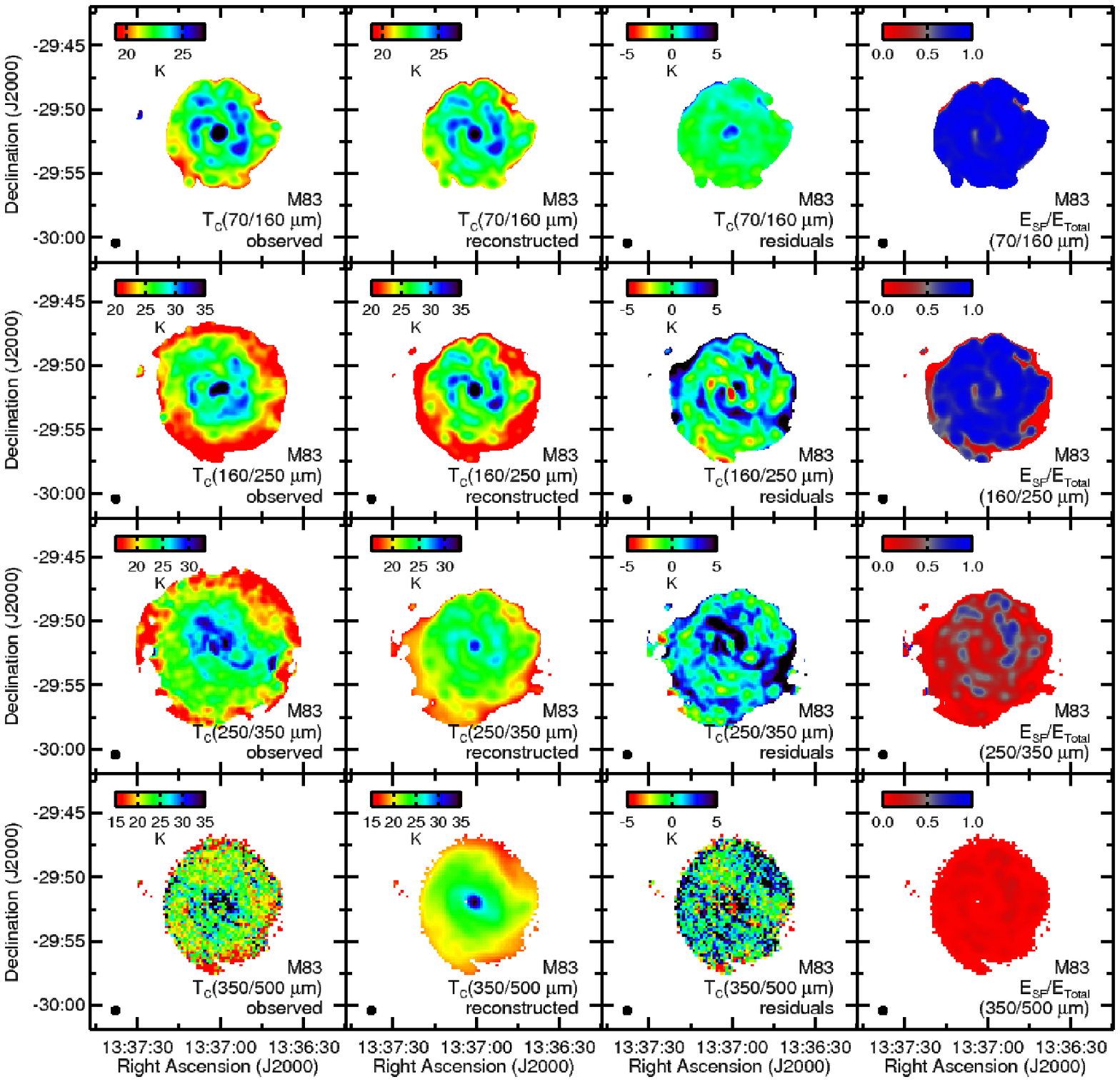}
\end{center}
\caption{The observed and reconstructed colour maps for M83, the
  residual values when the reconstructed maps are subtracted from the
  observed maps, and the $E_{SF}/E_{Total}$ maps.  Each map is
  $20\times20$~arcmin.  See the caption of
  Figure~\ref{f_mapfrac_ngc3031} for other information about the
  layout.}
\label{f_mapfrac_ngc5236}
\end{figure*}

\begin{figure*}
\begin{center}
\epsfig{file=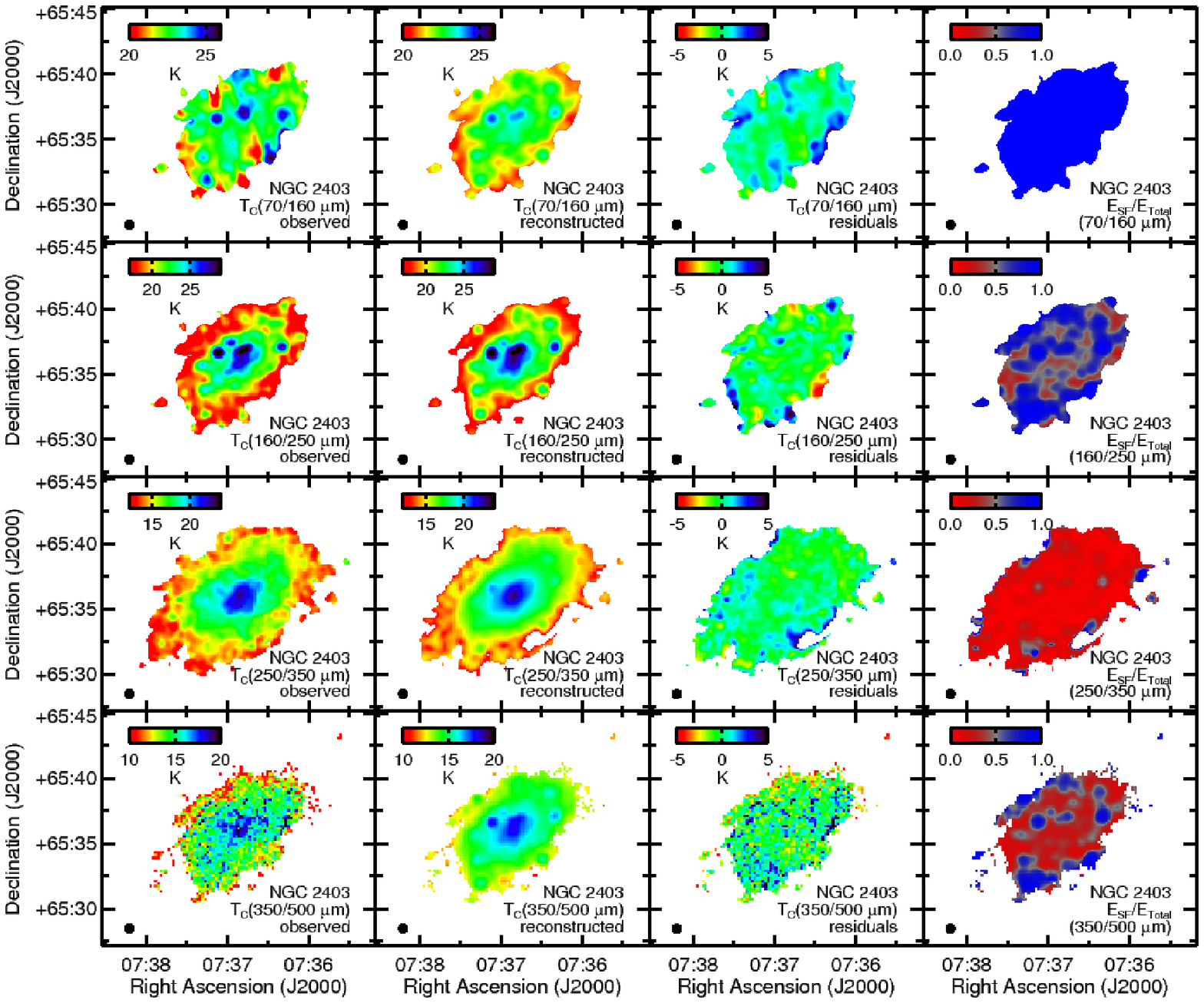}
\end{center}
\caption{The observed and reconstructed colour maps for NGC~2403, the
  residual values when the reconstructed maps are subtracted from the
  observed maps, and the $E_{SF}/E_{Total}$ maps.  Each map is
  $21\times18$~arcmin.  See the caption of
  Figure~\ref{f_mapfrac_ngc3031} for other information about the
  layout.}
\label{f_mapfrac_ngc2403}
\end{figure*}

We fit Equation~\ref{e_efit} to the 36~arcsec square subregions in
each galaxy, as it was easier to use this equation with a
Levenberg-Marquardt algorithm to converge upon the best fit to the
data, even though the solution is degenerate.  We then used these
results to calculate the parameters for Equation~\ref{e_efinal}.
Table~\ref{t_colorvsmult} given the best fitting parameters along with
the correlation coefficients for the values given by the right and
left sides of Equation~\ref{e_efinal} and the fractional contribution
of star forming regions to dust heating as given by
Equation~\ref{e_efrac}.  Uncertainties in the quantities in
Table~\ref{t_colorvsmult} are calculated using a monte carlo approach
in which we perform fits in a series of iterations by adding Gaussian
noise to the data and then measure the standard deviation in the
results. Figure~\ref{f_multfit} shows the relations between the
surface brightness ratios and the best fitting relations.
Additionally, we used these results along with Equation~\ref{e_ct} to
create colour temperature maps constructed from the H$\alpha$ and
1.6~$\mu$m images.  These reconstructed colour temperature maps as
well as the colour temperature maps based on the observed surface
brightness ratios and maps of the relative contribution of star
forming regions to dust heating are all shown in
Figures~\ref{f_mapfrac_ngc3031}-\ref{f_mapfrac_ngc2403}.

\begin{table*}
\centering
\begin{minipage}{154mm}
\caption{Results from fitting surface brightness ratios as a function 
    of both H$\alpha$ and 1.6~$\mu$m emission.}
\label{t_colorvsmult}
\begin{tabular}{@{}lcccccc@{}}
\hline
Galaxy &      Surface &             
    $A_1^{ab}$ &        $A_2^a$ &        $\alpha^a$ &
    $E_{SF}/E_{Total}~^c$ &      
    Correlation\\
&             Brightness Ratio &    
    &                 
    &
    &  
    &  
    Coefficient$^d$\\
\hline
M81 &        70/160~$\mu$m &          
    $(2.25 +/- 0.05)\times10^{-12} $ &       
    $12.01 \pm 0.12$ &            
    $0.369 \pm 0.004$ &            
    $0.361 \pm 0.006$ &   
    0.76\\
&            160/250~$\mu$m &         
    $(5.4 +/- 0.6)\times10^{-12} $ &       
    $7.3 \pm 0.2$ &            
    $0.185 \pm 0.005$ &            
    $0.19 \pm 0.02$ &   
    0.89\\
&            250/350~$\mu$m &         
    $(1.3 +/- 0.4)\times10^{-11} $ &       
    $4.3 \pm 0.2$ &            
    $0.100 \pm 0.006$ &            
    $0.09 \pm 0.03$ &   
    0.93\\
&            350/500~$\mu$m &         
    $(1.2 +/- 1.1)\times10^{-11} $ &       
    $3.6 \pm 0.3$ &            
    $0.078 \pm 0.007$ &            
    $0.10 \pm 0.06$ &   
    0.82\\
M83 &        70/160~$\mu$m &          
    $(5.7 +/- 1.2)\times10^{-13e}$ &       
    $8.91 \pm 0.12^e$ &            
    $0.274 \pm 0.004^e$ &            
    $0.924 \pm 0.014^e$ &   
    $0.89^e$\\
&            160/250~$\mu$m &         
    $(1.06 +/- 0.08)\times10^{-12} $ &       
    $5.21 \pm 0.02$ &            
    $0.120 \pm 0.001$ &            
    $0.854 \pm 0.009$ &   
    0.88\\
&            250/350~$\mu$m &         
    $(1.02 +/- 0.17)\times10^{-11} $ &       
    $2.24 \pm 0.03$ &            
    $0.038 \pm 0.001$ &            
    $0.38 \pm 0.04$ &   
    0.85\\
&            350/500~$\mu$m &         
    $(1 +/- 3)\times10^{-10} $ &       
    $2.63 \pm 0.06$ &            
    $0.048 \pm 0.001$ &            
    $0.06 \pm 0.04$ &   
    0.66\\
NGC 2403 &   70/160~$\mu$m &          
    $0^f$ &       
    $4.23 \pm 0.02^f$ &            
    $0.144 \pm 0.002^f$ &            
    $1^f$ &   
    0.51$^f$\\
&            160/250~$\mu$m &         
    $(6.2 +/- 0.6)\times10^{-12} $ &       
    $6.39 \pm 0.10$ &            
    $0.158 \pm 0.003$ &            
    $0.65 \pm 0.02$ &   
    0.89\\
&            250/350~$\mu$m &         
    $(1.1 +/- 0.3)\times10^{-10} $ &       
    $3.32 \pm 0.09$ &            
    $0.077 \pm 0.002$ &            
    $0.10 \pm 0.03$ &   
    0.93\\
&            350/500~$\mu$m &         
    $(1.6 +/- 0.6)\times10^{-11} $ &       
    $2.83 \pm 0.13$ &            
    $0.055 \pm 0.002$ &            
    $0.42 \pm 0.08$ &   
    0.76\\
\hline
\end{tabular}
$^a$ These parameters correspond to the ones in Equation~\ref{e_efinal}.\\
$^b$ The units of $A_1$ are (erg cm$^{-2}$ s$^{-1}$ arcsec$^{-2}$)
     (MJy sr$^{-1}$)$^{-1}$.\\
$^c$ This is relative fraction of dust heating from star forming
     regions as given by Equation~\ref{e_efrac}.  The value is
     calculated with the data for all 36~arcsec square regions used
     when fitting Equation~\ref{e_efinal} to the data.\\
$^d$ These are correlation coefficients for the relations between the 
     right and left sides of Equation~\ref{e_efinal}.\\
$^e$ These fits were performed excluding the central $3\times3$ 36~arcsec
     square regions.  See the text for details.\\
$^f$ Instead of reporting the result from the fit
     to Equation~\ref{e_efit} here, we instead report the result of the
     fit of a power law relation between the 70/160~$\mu$m surface
     brightness ratio and the H$\alpha$ intensity.  See the text for details.\\
\end{minipage}
\end{table*}

We encountered two problems with fitting Equation~\ref{e_efit} to the
data.  In fitting the 70/160~$\mu$m surface brightness data for M83,
we found that the difference between the ratio and the resulting fit
was significantly high around the nucleus, and the reconstructed colour
temperature map in Figure~\ref{f_mapfrac_ngc5236} did not look like
the observed colour temperature map.  Additionally, the fractional
contribution of star formation to dust heating seemed abnormally low;
when the nucleus is included, the fractional contribution of star
formation to dust heating is $\sim0.3$.  This implies that the nucleus
does not follow the same 70/160~$\mu$m dust heating relation as the
rest of the disc, for reasons that are unclear but that could be
related to hot ($\sim100$~K) dust heated by the nuclear starburst that
is visible in the 70~$\mu$m band.  Therefore, we
excluded the central $3\times3$ regions ($108\times108$~arcsec) of M83
for the analysis on the 70/160~$\mu$m ratio.  (Excluding the same
central region when fitting Equation~\ref{e_efit} to the 160/250,
250/350, or 350/500~$\mu$m data did not affect the fits signficantly.)
In fitting Equation~\ref{e_efit} to the 70/160~$\mu$m data for
NGC~2403, we found that the $C_3 I_\nu (1.6 \mu \mbox{m})$ term was
negligible compared to the $C_2 I(\mbox{H} \alpha)$ term, which caused
convergance problems when trying to perform a nonlinear least squares
fit.  We therefore report results in which the 70/160~$\mu$m ratio is
a function of only the H$\alpha$ intensities (which is consistent with
the interpretation that variations in dust heating traced by this
ratio depend primarily on star formation).

After dealing with these two exceptions, the resulting fits are
generally very good.  The correlation coefficients in
Table~\ref{t_colorvsmult} for the relations between the logarithm of
the surface brightness ratios and the expression on the right side of
Equation~\ref{e_efinal} are generally higher than the corresponding
correlation coefficients in either Tables~\ref{t_colorvsstar} or
\ref{t_colorvsha} for fits to either the 1.6~$\mu$m or H$\alpha$
emission, although in many cases, the improvement in the correlation
coefficients is marginal.  Moreover, most correlation coefficients in
Table~\ref{t_colorvsmult} are above 0.7, indicating that at least 50\%
of the variance in the data can be accounted for by the relations.
The only exceptions are the 70/160~$\mu$m ratio for NGC~2403, which we
identified as problematic up above, and the 350/500~$\mu$m ratio for
M83, which is very noisy in general.

The reconstructed colour temperature maps based on
Equation~\ref{e_efinal} in
Figures~\ref{f_mapfrac_ngc3031}-\ref{f_mapfrac_ngc2403} look
qualitatively similar to the colour temperature maps based on the
observed surface brightness ratios.  The reconstructed maps generally
match the overall radial gradients in temperature seen in many of the
observed maps, and the reconstructed maps also show that we could
reproduce the colour temperatures of many of the individual star
forming regions within the spiral structures of each galaxy.  However,
the temperatures of the reconstructed star forming regions do not
always precisely match the colour temperatures in the observed map.
This could be because of variable extinction among individual star
forming regions, which could cause the relation between star formation
(as traced by H$\alpha$ emission) and dust heating to vary.  We also
occasionally see some residual structures in temperature that show
features that are not quite described by the reconstructed map.  In
M81, the nucleus (excluded from the fits) appears notably cold in the
residual maps because of the synchrotron emission from the AGN.  The
nucleus of M83, which caused problems in the fits and was excluded,
appears as a hot spot in the residual 70/160~$\mu$m map.  Residual
spiral structure is seen in the 160/250~$\mu$m and 250/350~$\mu$m
residual maps, which could be indicative of dust heating gradients
across the spiral arms that are inadequately described by
Equation~\ref{e_efinal} and show the limitations of using this
approach to reproducing dust colour temperature variations.  Foyle et
al. (in preparation) found a similar offset between dust temperatures
and star formation in the spiral arms.  Finally, we see residual
structure in the 70/160~$\mu$m map of NGC~2403 that is clearly a
result of latent image effects in the 70~$\mu$m data.

The main purpose of these fits was to determine the relative
contributions of star forming regions and the total stellar
populations to dust heating.  The results in Table~\ref{t_colorvsmult}
generally show a decrease in the relative contribution of dust heating
from star formation as wavelength increases.  In M83 and NGC 2403,
star forming regions contribute $>50$\% of the total heating of the
dust traced by the $<250$~$\mu$m bands, although in M81, the dust
heating up to 70~$\mu$m appears to still be dominated by the total
stellar population.  At longer wavelengths, the relative contribution
of star forming regions decreases signficantly.  Star forming regions
appear to contribute only $<30$\% of the energy for heating the dust
traced by the 250/350~$\mu$m and 350/500~$\mu$m bands in all three galaxies.
The relation for the 350/500~$\mu$m ratio for NGC~2403 seems anomalous
in that it implies that $\sim40$\% of the energy for the dust heating
originates from star forming regions, which is higher than what is
implied by the fit to the 250/350~$\mu$m ratio for that galaxy.
However, we suspect that this is because the 350/500~$\mu$m data are
on the Rayleigh-Jeans side of the SED and are strongly affected by
noise.  Hence, colour variations related to dust heating are not
detected at a high signal-to-noise level in NGC~2403, and so the
methods we are using may not be effective in this one case.

Overall, the results here are consistent with the interpretations we
have provided for the results in Sections~\ref{s_analysis_stars} and
\ref{s_analysis_sf}.  At wavelengths shorter than 160~$\mu$m, most of
the dust appears to be heated by star forming regions in these
galaxies, which is why the 70/160~$\mu$m ratios appear more closely
dependent on the H$\alpha$ surface brightnesses.  At wavelengths
longer than 250~$\mu$m, the dust appears to be primarily heated by the
total stellar populations, and so the 250/350~$\mu$m and 350/500~$\mu$m
ratios are more strongly related to the 1.6~$\mu$m surface brightness.

\section{Discussion}
\label{s_discussion}

The results here have demonstrated that the total stellar population,
including evolved stars and not just stars in star forming regions,
plays a significant role in heating dust in nearby galaxies.  The
fraction of observed dust emission that is heated by the total stellar
population varies as a function of the wavelength.  While star forming
regions are the dominant heating source for dust observed at
$<160$~$\mu$m, the total stellar population, including evolved stars,
become the dominant heating source for the dust observed at
$>250$~$\mu$m.  We have been able to demonstrate that dust is heated
by the total stellar populations in spite of the fact that the dust
emission itself tends to be correlated with H$\alpha$ emission, which
should bias our results towards finding dust heated by star formation.
Moreover, we have shown that significant dust heating by the total
stellar population can be seen even in late-type spiral galaxies.

The results here validate the original conclusions drawn from IRAS
data by \citet{lh87}, \citet{ws87}, \citet{st92}, \citet{wg96}, and
\citet{kcbf04} as well as the conclusions drawn from {\it Spitzer}
data by \citet{hetal04} and \citet{cetal10} that evolved stellar
populations play an increasingly important role in heating dust
towards longer infrared wavelengths.  Additionally, the differences
between the results for M81, where we were able to demonstrate that
the bulge stars were heating dust even at 70~$\mu$m, and the results
for NGC~2403, where the dust emitting at 70~$\mu$m did not appear to
be influenced at all by the bulge stars, corroborates results from
\citet{eetal10}, who found that bulges influenced dust heating in
nearby galaxies.

Furthermore, the results here corroborate the findings from radiative
transfer models \citep[e.g.][]{dce08, b08, ptdfkm11} as well as other
dust emission models \citep[e.g.][]{detal07} that had suggested that
dust absoprtion and emission can only be accurately modelled if it
contains a dust component heated by star formation and a dust
component heated by the total stellar population.  Both our results
and the radiative transfer analyses suggest that the $<160$~$\mu$m
emission originates from dust that is heated by star forming regions
located within clouds (or clumps) that are optically thick at
ultraviolet and blue wavelengths.  In these models, only a relatively
small fraction of the dust is heated by the star forming regions, but
the dust appears significantly brighter than the dust outside the
centres of the regions, leading to higher 70/160~$\mu$m surface
brightness ratios in these regions.  Most of the dust mass in the ISM,
however, is shielded from the ultraviolet and blue photons from the
star forming regions.  This dust is heated by the diffuse interstellar
radiation field, which would include light from evolved stars, and has
lower temperatures.  None the less, because a much greater fraction of
the dust mass is heated this way, the colder dust is the predominant
source of emission at $>250$~$\mu$m.  Also, if the mean free path of
light in the diffuse interstellar medium is similar to or smaller than
the resolution elements in our analysis, the 250/350~$\mu$m and
350/500~$\mu$m ratios can be expected to be very well correlated with
the stellar surface brightnesses in those resolution elements.  The
V-band optical depths and the dust vertical scale heights derived in
the analyses of edge-on galaxies by \citet{xbkpp99} and \citet{b07}
imply that the mean free paths of V-band photons through the discs of
spiral galaxies should be roughly between 0.1 and 1~kpc.  In
comparison, the binned data that we used correspond to physical sizes
of $\sim500$-800~pc, and so the 250/350~$\mu$m and 350/500~$\mu$m ratios
should be very well correlated with tracers of the total stellar
population in these binned data, which is exactly what we see.

While we do produce observational results that are generally
consistent with many previously-published dust models, we do find a
inconsistency when comparing our results for M81 and NGC~2403 with the
dust models for these specific galaxies from \citet{detal07}.
\citet{detal07} did predict from their SED fitting to M81 that most of
the dust emission at $>70$~$\mu$m is from a diffuse component, which
is largely consistent with the results from our surface brightness
ratio variations.  However, their SED fitting to NGC~2403 also
predicted that most of the dust emission at $>70$~$\mu$m from that
galaxy was from diffuse dust, whereas the surface brightness
variations we observed imply that the star forming regions should be
the predominant heating source at $<160$~$\mu$m.  This demonstrates
that, while the general ideas being applied in dust models are still
valid, further refinement is needed if the models are not only going
to replicate the overall SEDs of galaxies but also replicate colour
variations within the galaxies.

Even though many authors have argued that a significant fraction of
the dust in nearby spiral galaxies is heated by evolved stellar
populations, other authors have shown that total infrared dust
emission can be correlated with star formation in nearby spiral
galaxies \citep{dy90, djc95, bx96, ketal09}, and even some of the
early results from {\it Herschel} suggest that emission in individual
wave bands (i.e. the 100, 160, and 250~$\mu$m bands) can be correlated
with other star formation tracers in M33 \citep{bcketal10, vetal10}, a
galaxy with the same Hubble type as NGC~2403.  We think the apparent
contradictions between the correlations between far-infrared emission
and star formation found by other authors and the relation between
dust heating and the total stellar populations that we find can be
explained in two ways.

First of all, some of the correlations between far-infrared emission
and star formation are based on either the far-infrared fluxes traced
by the IRAS 60-100~$\mu$m data or on the total dust emission as
computed from the {\it Spitzer} 24-160~$\mu$m bands, both of which
strongly sample emission from dust on the Wien side of the thermal
dust emission in a regime where we also found that star formation
strongly influenced dust heating.  Moreover, IRAS-based studies simply
did not have data at $>$100~$\mu$m wavelengths, which we found were
dominated by emission from dust heated by the total stellar
population.

Second, it is possible that far-infrared emission could be linked to
star formation indirectly through the Schmidt law \citep{s59, k98b}
rather than directly through dust heating.  Dust emission at
far-infrared wavelengths is a function of both dust temperature and
dust surface density.  Since dust should implicitly be a tracer of the
total surface density of gas in the interstellar medium, and since the
Schmidt law suggests that star formation is a function of the amount
of gas available to fuel it, we expect dust emission to be correlated
with star formation to some degree even if the dust is not directly
heated by the star formation.  The dust colours, however, are strongly
affected by dust temperature but are independent of dust surface
density and therefore will depend more on dust heating sources than
the emission observed in a single wave band.  This is probably why we
are able to show that far-infrared surface brightness in the
250-500~$\mu$m bands is correlated with H$\alpha$ intensity even
though we have demonstrated that the 250/350~$\mu$m and 350/500~$\mu$m
surface brightness ratios are more strongly correlated with 1.6~$\mu$m
surface brightness.  Additionally, in the case of M81, the 1.6~$\mu$m
emission is poorly correlated with the 250, 350, and 500~$\mu$m
surface brightness, and yet the 250/350~$\mu$m and 350/500~$\mu$m
ratios are clearly being heated mostly by the stellar population
traced by the 1.6~$\mu$m band.  In this case the infrared surface
brightness can only be correlated with the H$\alpha$ emission if the
dust surface density is correlated with the star formation surface
density.  We believe that this may also be a possible reason why
\citet{bcketal10} and \citet{vetal10} can show that the 100-250~$\mu$m
emission is correlated with H$\alpha$ and 24~$\mu$m emission, although
a comparison of the infrared colour variations in M33 to tracers of
star formation or the total stellar population would be needed to
confirm this.

Aside from the use of surface brightness ratios instead of surface
brightnesses or flux densities in our analysis, we think that the
choice of galaxies that we used in this analysis was critical to
demonstrating that the total stellar populations (including evolved
stars) play a significant role in dust heating.  M81 and NGC~2403 are
both special in that the stellar emission traces different structures
than either the dust emission or the star formation.  This has been
critically important for allowing us to disentangle the relative
contributions from the two different heating sources.  M83 is a good
example of a ``typical'' spiral galaxy in which the starlight, dust
emission, and star formation all trace similar structures.  Our
analysis demonstrates that it is difficult to disentangle whether the
total stellar population or star formation is a greater contributor to
dust heating in such cases.  Other authors working with M83 or similar
galaxies will probably encounter similar problems.

We emphasize that we have only found that these results apply to
spiral galaxies.  \citet{gmgetal10} demonstrated that the
250/500~$\mu$m surface brightness ratio depends on 24~$\mu$m surface
brightness in the dwarf irregular galaxy NGC~6822, thus demonstrating
that the dust emitting at submillimetre wavelengths was significantly
affected by star forming regions.  This is probably a result of the
galaxy containing relatively few evolved stars that could heat dust,
although it is conceivable that, in the low metallicity environment,
the interstellar medium would contain less dust, and so diffuse dust
would be less shielded from star forming regions.  We anticipate that
similar results would be obtained for other dwarf irregular galaxies
with ongoing star formation.  In contrast, we may expect that the
$>$70~$\mu$m emission seen in elliptical galaxies (when the elliptical
galaxies produce any far-infrared emission at all) will originate from
dust predominantly heated by evolved stars, just as the dust from the
central 3~kpc of M81 appears to be heated primarily by evolved stars.
How our results for spiral galaxies apply to luminous infrared
galaxies (with total infrared luminosities between $10^{11}$ and
$10^{12}$ L$_\odot$) and ultraluminous galaxies (with total infrared
luminosities above $10^{12}$ L$_\odot$) is unclear, as the galaxies
contain strong star formation and AGN emission but also have large
stellar populations and may have large reservoirs of dust that is
shielded from the star forming and AGN regions.  Additional analyses
on the dust heating mechanisms in these classes of galaxies is
warranted.

\section{Conclusions}
\label{s_conclusions}

For M81, M83, and NGC~2403, we have demonstrated that the
250/350~$\mu$m and 350/500~$\mu$m surface brightness ratios are more
strongly correlated with 1.6~$\mu$m emission, which traces the
starlight from the total stellar populations including evolved stars,
than with H$\alpha$ emission from star-forming regions.  At shorter
wavelengths, the total stellar population may continue to be
influential in dust heating, particularly in early-type spiral galaxy
M81, but heating by star formation becomes more important,
particularly in the late-type spiral galaxy NGC~2403.  None the less,
these results imply that, in each of these galaxies as well as other
spiral galaxies in general, the total stellar populations are a
significant if not dominant heating source for the dust observed at
$>$160~$\mu$m in all spiral galaxies, including late-type spiral
galaxies.  In some cases, virtually all of the emission may originate
from dust heated by the total stellar populations, including evolved
stars in the bulges and discs.

The results here have strong implications for modeling dust in spiral
galaxies.  First of all, dust models need to include separate thermal
components for the dust heated by star forming regions and dust heated
by the total stellar populations.  When fitting single modified
blackbodies to data, it may be inappropriate to force the function to
fit 70~$\mu$m or shorter wavelength data, particularly by varying the
index of the power law that describes the dust emissivity (see, for
example, \citet{detal00}, but also note the additional analysis in
\citet{de01}).  Models based on more complex dust physics need to
include components heated by red stellar radiation fields that only
become predominant at longer wavelengths, as is already done in some
cases, and these models also need to be tuned so that they replicate
not only the global SEDs of galaxies but also the colour variations
within galaxies.  Template-based SED models need to include an
additional component based on dust heated by a quiescent stellar
population, as is done by \citet{retal10}, although these cold dust
templates should be calibrated using observations of nearby galaxies
where the dust emission can be studied on kpc scales rather than
inferred from observations of unresolved sources.  Also, dust
extinction corrections that rely upon infrared flux to estimate
extinction corrections, as is commonly done for ultraviolet data
\citep[e.g. ][]{mhc99}, need to account for dust heating by evolved
stellar populations and not just star forming regions, as is also
suggested by \citet{kcbf04} and \citet{cbfdgbb08}.  Finally, we are
going to suggest caution when using far-infrared emission as a star
formation tracer for both nearby and more distant galaxies.  Dust
emission in individual wave bands at $>$160~$\mu$m is probably related
to star formation indirectly through the Schmidt law rather than
directly through dust heating.  Using, for example, 250~$\mu$m flux
density measurements to determine star formation rates is akin to
using CO measurements to determine star formation rates.  While this
may still be an accurate way to measure star formation rates under
certain circumstances, it will be necessary to understand the caveats
that affect such analyses.

\section*{Acknowledgments}

G.J.B. thanks Simone Bianchi and the reviewer for helpful comments on
this paper.  G.J.B. was funded by the STFC.  The reserach of
C.D.W. and K.F. is supported by grants to C.D.W. from the Canadian
Space Agency and the Natural Sciences and Engineering Research Council
of Canada.  PACS has been developed by a consortium of institutes led
by MPE (Germany) and including UVIE (Austria); KU Leuven, CSL, IMEC
(Belgium); CEA, LAM (France); MPIA (Germany); INAF-IFSI/OAA/OAP/OAT,
LENS, SISSA (Italy); IAC (Spain). This development has been supported
by the funding agencies BMVIT (Austria), ESA-PRODEX (Belgium),
CEA/CNES (France), DLR (Germany), ASI/INAF (Italy), and CICYT/MCYT
(Spain).  SPIRE has been developed by a consortium of institutes led
by Cardiff University (UK) and including Univ. Lethbridge (Canada);
NAOC (China); CEA, LAM (France); IFSI, Univ. Padua (Italy); IAC
(Spain); Stockholm Observatory (Sweden); Imperial College London, RAL,
UCL-MSSL, UKATC, Univ. Sussex (UK); and Caltech, JPL, NHSC,
Univ. Colorado (USA). This development has been supported by national
funding agencies: CSA (Canada); NAOC (China); CEA, CNES, CNRS
(France); ASI (Italy); MCINN (Spain); SNSB (Sweden); STFC (UK); and
NASA (USA).  HIPE is a joint development by the Herschel Science
Ground Segment Consortium, consisting of ESA, the NASA Herschel
Science Center, and the HIFI, PACS and SPIRE consortia.  This research
has made use of the NASA/IPAC Extragalactic Database (NED) which is
operated by the Jet Propulsion Laboratory, California Institute of
Technology, under contract with the National Aeronautics and Space
Administration.

{}

\label{lastpage}

\end{document}